\documentclass[aps,prd,twocolumn,floatfix,nofootinbib,showpacs,superscriptaddress,tightenlines]{revtex4-1}
\usepackage{amsmath}
\usepackage{lipsum}
\usepackage{amssymb}
\usepackage{amsthm}
\usepackage{bbold}
\usepackage{dcolumn}
\usepackage{epsfig}
\usepackage{graphics}
\usepackage{graphicx}
\usepackage{longtable}
\usepackage{color}
\usepackage{epstopdf}
\usepackage{xspace}
\usepackage{cancel}
\usepackage{nicefrac}
\usepackage[colorlinks=true, pdfstartview=FitV, linkcolor=purple, citecolor= purple,urlcolor=blue]{hyperref}

\definecolor{darkgreen}{rgb}{0,0.5,0}
\definecolor{purple}{rgb}{0.5,0,0.5}
\definecolor{nblue}{rgb}{0.0,0.0,0.50}
\definecolor{scarlet}{rgb}{1.0,0.2,0}

\newcommand{\M}{\scriptsize{\hbox{M}}}
\em
\begin{document}

\title{Electromagnetic and two-photon transition form factors of the pseudoscalar mesons: An algebraic model computation}

\author{I. M. Higuera-Angulo}
\email{melany.higuera@umich.mx}
\affiliation{Instituto de F\'isica y Matem\'aticas, Universidad
Michoacana de San Nicol\'as de Hidalgo, Morelia, Michoac\'an
58040, M\'exico}
\author{R. J. Hern\'andez-Pinto}
\email{roger@uas.edu.mx}
\affiliation{Facultad de Ciencias F\'isico-Matem\'aticas, Universidad Aut\'onoma de Sinaloa, Ciudad Universitaria, Culiac\'an, Sinaloa 80000, M\'exico}
\author{K. Raya}
\email{khepani.raya@dci.uhu.es}
\affiliation{Department of Integrated Sciences and Center for Advanced Studies in Physics, Mathematics and Computation, University of Huelva, E-21071 Huelva, Spain}
\author{A. Bashir}
\email{adnan.bashir@umich.mx}
\affiliation{Instituto de F\'isica y Matem\'aticas, Universidad
Michoacana de San Nicol\'as de Hidalgo, Morelia, Michoac\'an
58040, M\'exico}
\affiliation{Department of Integrated Sciences and Center for Advanced Studies in Physics, Mathematics and Computation, University of Huelva, E-21071 Huelva, Spain}

\date{\today}

\begin{abstract}
We compute electromagnetic and two-photon transition form factors
%, and corresponding charge radii, 
of ground-state pseudoscalar mesons: $\pi,\,K,\,\eta_c,\,\eta_b$. To this end, we employ an algebraic model based upon the coupled formalism of Schwinger-Dyson and Bethe-Salpeter equations. Within this approach, the dressed quark propagator and the relevant Bethe-Salpeter amplitude encode the internal structure of the corresponding meson. Electromagnetic properties of the meson are probed via the quark-photon interaction. The algebraic model employed by us unifies the treatment of all ground-state pseudoscalar mesons. Its parameters are carefully fitted performing a global analysis of existing experimental data including the knowledge of the charge radii of the mesons studied. We then compute and predict electromagnetic and two-photon transition form factors for a wide range of probing photon momentum-squared  which is of direct relevance to the experimental observations carried out thus far or planned at different hadron physics facilities such as the Thomas Jefferson National Accelerator Facility (JLab) and the forthcoming Electron-Ion Collider. We also present comparisons with other theoretical models and approaches and lattice quantum chromodynamics.
\end{abstract}

%\pacs{12.38.-t, 11.15.Tk, 14.40.Be, 14.40.Pq, 13.40.Gp}
%\keywords{}

\maketitle
\date{\today}

%%%%%%%%%%%%%%%%%%%%%%%%%%%%%%%%%%%%%%%%%%%%%%%%%%%%%%%%%%%%%%%
%%%%%%%%%%%%%%%%%%%%%%%%%%%%%%%%%%%%%%%%%%%%%%%%%%%%%%%%%%%%%%%
%%%%%%%%%%%%%%%%%%%%      INTRODUCTION     %%%%%%%%%%%%%%%%%%%%
%%%%%%%%%%%%%%%%%%%%%%%%%%%%%%%%%%%%%%%%%%%%%%%%%%%%%%%%%%%%%%%
%%%%%%%%%%%%%%%%%%%%%%%%%%%%%%%%%%%%%%%%%%%%%%%%%%%%%%%%%%%%%%%
\section{Introduction}
\label{Section Introduction}

Unraveling the internal structure of hadrons is one of the most intriguing, challenging and yet highly researched topics within the strong interaction physics. Although the underlying fundamental field theory, {\em i.e.}, quantum chromodynamics (QCD), was proposed on firm grounds some five decades ago, its most compelling success 
is predominantly restricted to its perturbative domain alone. The emergent non-perturbative phenomena of confinement, hadron masses and structural properties of even the simplest strongly interacting bound states are notoriously hard to be elucidated from the underlying first principles. The plain reason is that the well-established and well-trusted orthodox perturbative methods are no longer suitable for this undertaking~\cite{Gross:2022hyw,Accardi:2023chb}. 
Different non-perturbative approaches are relied upon, including: lattice QCD, coupled formalism of Schwinger-Dyson (SDEs) and Bethe-Salpeter equations (BSEs), and other effective theories and models (see, \emph{e.g.}~\cite{Constantinou:2020hdm,Bashir:2012fs,Aguilar:2019teb,PhysRevD.46.4052,Brambilla:2014jmp}).
From the experimental point of view, unveiling the properties of hadrons in terms of cross sections is equally challenging but essential to make progress through direct comparison and contrast with theory. Ongoing and planned facilities aim at probing the internal structure of hadrons at different energy scales with an unprecedented resolution and precision~\cite{Accardi:2023chb,Aznauryan:2012ba,Chavez:2021koz,Chen:2020ijn,Huber:2006}. 

Much like the hydrogen atom for quantum electrodynamics (QED), pions and kaons, the simplest two-body strongly interacting bound states, serve as a portal to understand the intricate dynamics of QCD.
They hold a unique relationship with the dynamical mechanism underlying the Emergent Hadronic Mass (EHM)~\cite{Arrington:2021biu,Horn:2016rip}. Comparison with the heavy counterparts of these light mesons, such as $\eta_{c}$ and $\eta_b$, reveals essential differences with the diametrically opposed mass generating mechanisms, namely, EHM and the mass triggered by the famous Higgs mechanism. These competing sources expose themselves via different hadron observables such as form factors and parton distributions. In this article, we focus on the form factors of ground-state pseudoscalar mesons ($\M=\pi,\,K,\,\eta_c,\,\eta_b$), including  electromagnetic ($\gamma^*\M  \to \M$) and transition ($\gamma\gamma^*\to\M$) form factors (labeled as EFF and TFF, respectively).

EFFs provide valuable information about the internal dynamics of hadrons, such as the charge and current distribution, and how asymptotic QCD predictions can be approached for sufficiently large photon momentum squared $Q^2$. It's of course more and more unlikely to prevent a meson from breaking up at large $Q^2$ and it becomes increasingly hard for the experiment to probe widely disparate momentum scales in one single experiment. Both the Thomas Jefferson National Accelerator Facility (JLab) and the planned Electron-Ion Collider (EIC) will push the observable range of $Q^2$ to unprecedentedly large intervals. On the other hand, TFFs
 $\gamma\gamma^*\to\M$ can be accessed to much larger $Q^2$ with relative ease as they rely on mesons production mechanism rather than keeping them intact; in this scenario, the main challenge would be primarily controlling the background produced by other processes.  
 
 Noticeably, pseudoscalar meson's EFF and two-photon TFF are characterized by a single form factor, hence facilitating its experimental extraction~\cite{Horn:2016rip}. Furthermore, these form factors provide a neat platform to test fundamental predictions of QCD, such as scaling violations and factorization formulae~\cite{Lepage:1980fj,Efremov:1979qk,Lepage:1979zb}. The latter entails that, at sufficiently large energy scales ($Q^2  \gg \Lambda_{\text{QCD}}^2$), asymptotic QCD limits are approached:
\begin{eqnarray}   
Q^2F_{\M}(Q^2) &\sim&   [f_{\M} \tilde{\mathnormal{w}}_{\M}^q(Q^2)]^2 \alpha_s(Q^2)\;,
    \label{eq:defEFFf}\\
Q^2G_{\M}(Q^2) &\sim&  [f_{\M} \tilde{\mathnormal{w}}_{\M}^q(Q^2)]^1 \;,
    \label{eq:defTFF}
\end{eqnarray}
where $F_{\M}$ and $G_{\M}$ denote the EFF and TFF, respectively; $\alpha_s(Q^2)$ is the strong coupling constant at one loop, $f_{\M}$ is the pseudoscalar meson leptonic decay constant and
\begin{eqnarray}
    \tilde{\mathnormal{w}}_{\M}^q(Q^2) = \int_0^1 dx \frac{1}{x} \phi_{\M}^q(x,Q^2) \;.
\end{eqnarray}
Here $\phi_{\M}^q(x,Q^2)$ refers to the leading-twist parton distribution amplitude (PDA)~\cite{Chang:2013pq}. It is a close analogue of the quantum mechanical wave function of the meson whose asymptotic form adopts the form:
\begin{eqnarray}
    \phi_{\M}^q(x,Q^2\to \infty) \to \phi_{asy}(x) = 6x (1-x) \;.\label{eq:PDAasy}
\end{eqnarray}
Scaling violations are explicit in Eq.~\eqref{eq:defEFFf} due to the presence of $\alpha_s(Q^2)$; and, although the expression of the TFF, Eq.~\eqref{eq:defTFF}, does not plainly exhibit such scaling violations, the way in which this asymptotic result is approached for increasing $Q^2$ is implicitly  governed by these scaling violations~\cite{Raya:2015gva}. The factorization theorems also reveal that, asymptotically, the form factors would be weighted by the decay constant, which is a measure of dynamical chiral symmetry breaking~\cite{Sultan:2018tet}. For the TFF, this behavior is manifest in the opposite energy domain where the Abelian chiral 
anomaly entails~\cite{Adler:1969gk,Bell:1969ts,Adler:2004ih}:
\begin{eqnarray}
 \label{eq:Abelian}
    2f_{\pi}^0G_{\pi} (Q^2=0) = 1 \;,
\end{eqnarray}
where $f_{\pi}^0=0.092$ GeV is the value of the pion decay constant in the chiral limit. 

Several experimental efforts are aimed at measuring the pion EFF~\cite{AMENDOLIA1986168,ACKERMANN1978294,JeffersonLabFpi:2000nlc,JeffersonLabFpi-2:2006ysh,JeffersonLab:2008jve} as well as TFF~\cite{CELLO:1990klc,PhysRevD.57.33,PhysRevD.80.052002,PhysRevD.86.092007}. Notably, for the pion EFF, all collaborations converge to an excellent agreement within the available domain of results ({\em i.e.}, up to $2.5\,\text{GeV}^2$). Notwithstanding, current trend of experimental results falls  short of approaching the asymptotic limit, {\em i.e.}, Eq.~\eqref{eq:defEFFf}. JLab and EIC planned experiments aim to extend the upper bound of observed $Q^2$ to around 8.5 GeV$^2$ (12 GeV upgrade), 15 GeV$^2$ (potential 22 GeV upgrade), and 35 GeV$^2$~\cite{Arrington:2021biu,Aguilar:2019teb,Accardi:2023chb,Horn:2016rip,Huber:2006}, respectively.  

On the other hand, there remains an existing controversy concerning the two-photon pion TFF. In this case, while all available experimental data agree at low to moderate values of $Q^2$,  the results above $Q^2 \sim 10\,\text{GeV}^2$ of the BaBar collaboration,~\cite{PhysRevD.80.052002}, disagree markedly with those from the Belle collaboration~\cite{PhysRevD.86.092007} and from the perturbartive QCD prediction of Eq.~\eqref{eq:defTFF} for sufficiently large values of $Q^2$. Moreover, is appears to be at par with all other extant pion properties. This marked discrepancy has resulted in some analyses to argue that the BaBar data might not be considered an accurate measure of the pion TFF~\cite{Roberts:2010rn,Brodsky:2011yv,Bakulev:2011rp,Brodsky:2011xx,Stefanis:2012yw,Bakulev:2012nh,El-Bennich:2012mkr,Melikhov:2014gf,Raya:2015gva}. Concerning the kaon, empirical information is currently scarce and limited to the low-$Q^2$ regime~\cite{AMENDOLIA1986435,PhysRevLett.45.232,PhysRevC.97.025204}; but more data, and in a larger domain, which is crucially required, is under preliminary analysis,~\cite{JLabNew}. It is worth recalling that there is available BaBar data for the $\gamma\gamma^{\ast}\rightarrow \eta_c$ transition which extends up to probing photon momentum squared $Q^2>35\,\text{GeV}^2$, displaying satisfactory consistency with Eq.~\eqref{eq:defTFF},~\cite{Raya:2016yuj}. Note that no experimental results are yet reported for the transition $\gamma\gamma^{\ast} \hspace{-1mm} \rightarrow  \hspace{-1mm} \eta_b$ for which predictions are available~\cite{Raya:2016yuj}. 
 Within a set of sophisticated truncations of SDEs/BSEs, several form factors have been scrutinized, including $\gamma^* \pi^+ \hspace{-1mm} \to \hspace{-1mm} \pi^+$ and  $\gamma^* K^+ \hspace{-1mm} \to K^+$ EFFs~\cite{Chang:2013nia,Gao:2017mmp,Eichmann:2019bqf,Miramontes:2021exi,Xu:2023izo,Raya:2022ued}, TFFs $\gamma^* \gamma^* \to \{ \pi^0, \eta,\,\eta',\,\eta_c,\,\eta_b \}$~\cite{Raya:2015gva,Raya:2016yuj,Ding:2018xwy,Raya:2019dnh,EICHMANN2017425}. Time-like EFFs~\cite{Miramontes:2023ivz,Miramontes:2022uyi,Miramontes:2021xgn} have also begun to emerge which extends the domain of applicability of the SDE formalism. Then there are much simpler SDE-based  Contact Interaction (CI) results~\cite{Roberts:2010rn, article,Bedolla:2016yxq,Xing:2021dwe, Wang:2022mrh,Hernandez-Pinto:2023yin,Dang:2023ysl,Xing:2024bpj}, whose reliable applicability is limited to small $Q^2$ values. We work with a recently proposed Algebraic Model (AM) which captures large $Q^2$ properties of QCD fairly well while still preserving the simplicity 
 of the CI.
 
This manuscript is organized as follows: in Sec.~\ref{sec:formalism} we describe the basic elements of the AM employed throughout this work, and construction of the quark-photon vertex (QPV); in Sec.~\ref{sec:elasticff} we derive semi-analytical expressions for the EFF of pseudoscalar mesons, considering the most generic scenario, {\emph{i.e.}}, with different flavours of dressed quarks; Sec.~\ref{sec:transitionff} contains an analogous derivation for the TFFs of pseudoscalar mesons. Sec.~\ref{sec:numerics} details the results of our global fit of EFFs and TFFs within the AM, related to the best fitted parameters for pion, kaon, $\eta_c$ and $\eta_b$ mesons. Finally,  Sec.~\ref{sec:conclusions} is dedicated to the conclusions of this work and perspectives for future research directions.

\section{The formalism}
\label{sec:formalism}
The internal structure of the pseudoscalar meson is encoded within the so called Bethe-Salpeter amplitude (BSA), which in turn is related with the fully dressed quark propagator through the BSE and the axial vector Ward-Takahashi identity. Important aspects of the structural properties of the meson are exposed via its electromagnetic interaction with photons, such as those unravelled by the EFFs and TFFs. Within the present approach, the standard application of the Feynman rules to the meson-photon interactions, {\emph{i.e.}}, those describing the  $\M \gamma \M$ and $\gamma \M \gamma$ vertices, demands the knowledge of the QPV. Moreover, the complete evaluation of the form factors requires prior knowledge of the dressed quark propagator, the meson BSA, and the QPV. Their construction is discussed below.

\subsection{The algebraic model}

%The interaction description of the constituent quarks of a meson with photons at low energies is not computable from first principles. In this work, in order to surpass this difficulty, we present the analysis of EFF and TFF of pseudoscalar mesons in the context of an AM.

In order to systematically describe the structure of a pseudoscalar meson, $\M$, a typical starting point is the derivation of the Bethe-Salpeter wave function (BSWF). This is expressed in terms of the $q$-quark propagator $S_q$ and the corresponding BSA, $\Gamma_{\M}$, in the following form:
\begin{eqnarray}
    {\chi}_{\M} \left(k_{-}, P \right) = S_q(k) \Gamma_{\M}\left(k_{-}, P\right) S_{\bar{q}'}\left(k-P\right) \,,
\label{BS1}
\end{eqnarray}
where $q$ and $\bar{q}'$ correspond to the valence quark and antiquark respectively, $k_{-}=k-P/2$ ($k$ is the relative momentum between the two quarks), and $P^2=-m_{\M}^2$ is the squared mass of the meson. The quark propagator and BSA can be obtained by solving the corresponding SDE and BSE, which might require arduous work~\cite{Qin:2020rad}. However, taking advantage of the advances and understanding in this direction, it is possible to build simple, yet efficient and reliable {\em Ans$\ddot{a}$tze} of these entities that observe key QCD symmetries and mimic the expected behavior of the numerical solutions, and thus translate the full numerical approach to an amicable and trustable algebraic analysis in some measure. 

A family of such models have been employed in, among others, Refs.~\cite{Bedolla:2016yxq,Mezrag:2016hnp,Xu:2018eii,Raya:2021zrz,Albino:2022gzs,Almeida-Zamora:2023bqb} for pseudoscalar mesons and in Ref.~\cite{Almeida-Zamora:2023rwg} for the vector meson case. We adopt the phenomenologically motivated AM described in~\cite{Albino:2022gzs}, which is determined once the quark propagator and the pseudoscalar meson BSA are specified as follows:
\begin{eqnarray}
S_{q(\bar{q}')}(k) &=& \left[-i\gamma \cdot k + m_{q(\bar{q}')}\right] \Delta \left(k^2;m_{q(\bar{q}')}^2 \right)  \,,
\label{eq:Anzats1} \\
n_{\M} \Gamma_{\M}(k, p) &=& i\gamma_5 \int_{-1}^{1} dw \, \rho_{\M}(w)\left[ \hat{\Delta} \left(k_{w}^2 ; \Lambda_w^2 \right) \right]^\nu \hspace{-.1cm},
\label{eq:Anzats2}
\end{eqnarray}
where $\Delta(a;b)=(a+b)^{-1},\;\hat{\Delta}(a;b) = b \Delta(a;b)$ and $ k_{w} = k + (w/2)p$. Herein, $\nu=1+\delta$ is a parameter that controls the asymptotic behavior of the BSA, with $\delta$ playing the role of an anomalous dimension; $m_{q(\bar{q}')}$ is a mass scale that corresponds to the dressed mass for a given quark (antiquark) flavor $q(\bar{q}')$, and $n_{\M}$ is a normalization constant. Moreover,  $\rho_{\M}(w)$ is identified with a spectral density whose form dictates the pointwise behavior of the BSA and has  significant impact on a kaleidoscopic array of meson observables\footnote{We refer to the Appendix for the extraction of the spectral density function in the AM.}. As explained in~\cite{Albino:2022gzs} in great detail, the spectral density can be determined if the PDA is known. This process largely anticipates that an appropriate PDA would produce sensible form factors, distribution functions, generalized parton functions, and other related physical observables. 
 Finally, $\Lambda_w^2 \equiv \Lambda^2(w)$ is defined as follows:
\begin{eqnarray}
\hspace{-4mm} \Lambda_w^2 \hspace{-0.1cm} &\equiv & \hspace{-0.1cm} m_{q}^2-\frac{1}{4}\left(1-w^2\right)m_{\M}^2 +\frac{1}{2} \left( 1 -w  \right)\left(m_{\bar{q}'}^2-m_q^2\right). 
%\nonumber \\ 
\label{eq:Lambda}
\end{eqnarray}
It leads to the requirement that $m_{\M}<m_q+m_{\bar{q}'}$
and prevents the appearance of zeroes in the denominator. The model for the quark propagator and the BSA do not impose quark confinement. However, the constraint of Eq.~(\ref{eq:Lambda}) is naturally  satisfied by the Nambu-Goldstone bosons (\emph{i.e.} $\pi,\,K$). 
Moreover, numerical solutions for the heavier pseudoscalar mesons ($\eta_c,\,\eta_b$),\,\emph{e.g.}, Refs.~\cite{Sultan:2018tet,Ding:2018xwy} are also consistent with it. The spectral densities constructed through this {\em ansatz}  correspond to a non point-like state, and produce momentum-dependent form factors in agreement with the computation through other theoretical tools as well as experiment.

We would like to point out that with the simple structure required for the quark propagator and the BSA, Eq.~\eqref{eq:Anzats1} and Eq.~\eqref{eq:Anzats2}, the complete BSWF in the AM would be fully determined by the parameters $m^2_{q(\bar{q}')}$ %$m^2_{\M}$ 
and $\nu$. An additional one, $\alpha_q^{(0)}$, arising from the QPV, is introduced later on. It leaves us with a total of 3-4 parameters to be specified, depending on whether the quark and the antiquark are of the same flavor or observe isospin symmetry. Moreover, these parameters have little maneuverability 
as we have a good knowledge of the dressed quark masses, meson masses and the anomalous dimensions. The complete set of parameters shall be constrained through employing both experimental measurements and our theoretical understanding of the hadron structure. To carry out the calculation of these observables, what remains is the reliable knowledge of the interaction of the electromagnetic probe with the dressed quarks, {\emph{i.e.}}, the QPV; this is what we set out to discuss in the next subsection.

\subsection{The quark-photon vertex}
As stated before, the extraction of EFF and TFF is related with the $\M\gamma\M$ and $\gamma\M\gamma$ vertices, respectively. At the fundamental level, these meson-photon interactions are expressed in terms of the fully-dressed QPV, $\Gamma_\mu^q$. Hence, a proper construction of the latter becomes essential. It turns out that it is convenient to work with the unamputated version of the vertex, which reads:
\begin{eqnarray}
   \chi_{\mu}^q(k_f,k_i) = S_q(k_f)\Gamma_{\mu}^q(k_f,k_i)S_q(k_i) \,.
\end{eqnarray}
We thus adopt the  {\em Ansatz} described in Ref.~\cite{Raya:2015gva}:
\begin{eqnarray}
\label{eq:QPV}
    \chi_\mu^q(k_f,k_i) &=& T^{(1)}_\mu \Delta_{k^2 \sigma_v}+ T^{(2)}_\mu \Delta_{\sigma_v} + T^{(3)}_\mu \Delta_{\sigma_s}\,,\\
T^{(1)}_\mu &=& \gamma_{\mu} \,, \nonumber \\
T^{(2)}_\mu &=& \not\!{k_f}\gamma_{\mu} \not\!{k_i} + \alpha_q(\not\!{k_f}\gamma_{\mu} \not\!{k_i} - \not\!{k_i}\gamma_{\mu} \not\!{k_f}) \,, \nonumber \\
T^{(3)}_\mu &=& i(\not\!{k_f}\gamma_{\mu} + \gamma_{\mu}\not\!{k_i}) \nonumber \\
&+& i\alpha_{q} (\not\!{k_f}\gamma_{\mu} + \gamma_{\mu}\not\!{k_i} - \not\!{k_i}\gamma_{\mu} - \gamma_{\mu}\not\!{k_f}) \,;\nonumber
\end{eqnarray}
here $\Delta_F = [F(k_f^2)-F(k_i^2)]/[k_f^2-k_i^2]$ and $\sigma_{v,s}$ are the quark propagator dressing functions~\cite{Sultan:2018tet}:
\begin{equation}
     \sigma_s(p^2)=M(p^2) \sigma_v(p^2) = \frac{Z(p^2) M(p^2)}{p^2+M^2(p^2)}\;;\, 
\end{equation}
here $Z(p^2):=1$ as inferred from Eq.\,\eqref{eq:Anzats1}.
The part proportional to $\alpha_q$ is transverse to the external photon momentum $Q_\mu$. It is introduced as a momentum redistribution factor in the TFF, owing to the impossibility to simultaneously conserve the vector and axial-vector currents, which would make it prohibitively difficult to satisfy the Abelian anomaly of Eq.~\eqref{eq:Abelian}. As long as $\alpha_q$ is rapidly damped asymptotically, the point particle limit, $\Gamma_\mu \to \gamma_\mu$, is properly recovered. With that in mind, this factor is expressed as
\begin{equation}
\label{eq:alphaq}
    \alpha_q = \alpha^{(0)}_{q} \, \exp[-Q^2/(2m_q^2)] \,,
\end{equation}
where the parameter $\alpha^{(0)}_{q}$ sets the strength of the transverse pieces of the QPV.
The presence of $m_q$ as a flavor-dependent mass scale is natural, given that the transverse structures of the QPV are closely linked to the dynamical breaking of chiral symmetry\,\cite{Albino:2018ncl}.  The value $\alpha^{(0)}_{q}$ plays an essential role in ensuring the correct normalization of $G_{\M}(0)$, hence guaranteeing a proper description of the TFFs near vanishingly small momenta. The way the $\alpha_q$ term impacts the TFF, in fact, is similar to the contribution of the anomalous magnetic moment of the quark, whose importance for this and the anomalous $\gamma^* \to 3\pi$ process has recently been explained in Refs.~\cite{Dang:2023ysl,Xing:2024bpj}. Conversely, $F_{\M}(Q^2)$ and the associated charge radius are completely independent of $\alpha^{(0)}_{q}$; firstly, because the value $F_{\M}(Q^2 = 0) =1$ is completely fixed by the longitudinal part of the QPV and, secondly, due to the fact that only the dominant meson BSA has been considered,~\cite{Raya:2015gva}.

It is worth mentioning that the {\em Ansatz} for the QPV  in Eq.~\eqref{eq:QPV} has shown its robustness in the computation of the pion EFF~\cite{Chang:2013nia} and all two-photon TFFs involving ground-state neutral pseudoscalar mesons,~\cite{Raya:2015gva,Raya:2016yuj,Ding:2018xwy,Raya:2019dnh}. Furthermore, having been derived from the so-called gauge-technique~\cite{Delbourgo:1977jc}, the present vertex construction fulfills crucial mathematical requirements~\cite{Albino:2018ncl,Bermudez:2017bpx}: it is free of any kinematic singularities, the free-field limit is properly recovered, and it satisfies its corresponding Ward-Green-Takahashi identity. The latter identity is crucial in the present approach, since it enables us to express the QPV dressing functions in terms of those of the quark propagator, facilitating any related computation. Furthermore, given the simple character of the quark propagator in the present AM, Eq.~\eqref{eq:Anzats1}, the QPV can be systematically cast in a compact manner:
\begin{eqnarray}
\chi_{\mu}^q(k_f,k_i) = \frac{\sum_{j=1}^3 T^{(j)}_\mu X_j}{[k_f^2+m_q^2][k_i^2+m_q^2]} \,,
\label{eq:Quark-photon vertex}
\end{eqnarray}
where $X_1 = m_q^2 \,, \,\, X_2 = -1  \,, \,\, X_3 = -m_q$.

 Finally, it is worth recalling that this construction of the QPV introduces only one extra parameter to be determined, {\em {i.e.}}, $\alpha^{(0)}_{q}$. Next, in Sec.\,\ref{sec:elasticff} and Sec.\,\ref{sec:transitionff}, we proceed to derive the expressions related to the EFF and TFF, respectively. The corresponding numerical results are discussed in Sec.\,\ref{sec:numerics}.

%%%%%%%%%%%%%%%%%%%%%%%%%%%%%%%%%%%%%%%%%%%%%%%%%%%%%%%%%%%%%%%%%%%%%%%
%%%%%%%%%%%%%%%%%%%%%%%%%%%%%%%%%%%%%%%%%%%%%%%%%%%%%%%%%%%%%%%%%%%%%%%

\section{Electromagnetic form factors}
\label{sec:elasticff}
\begin{figure}[!h]
    \centering
    \includegraphics[scale=.31]{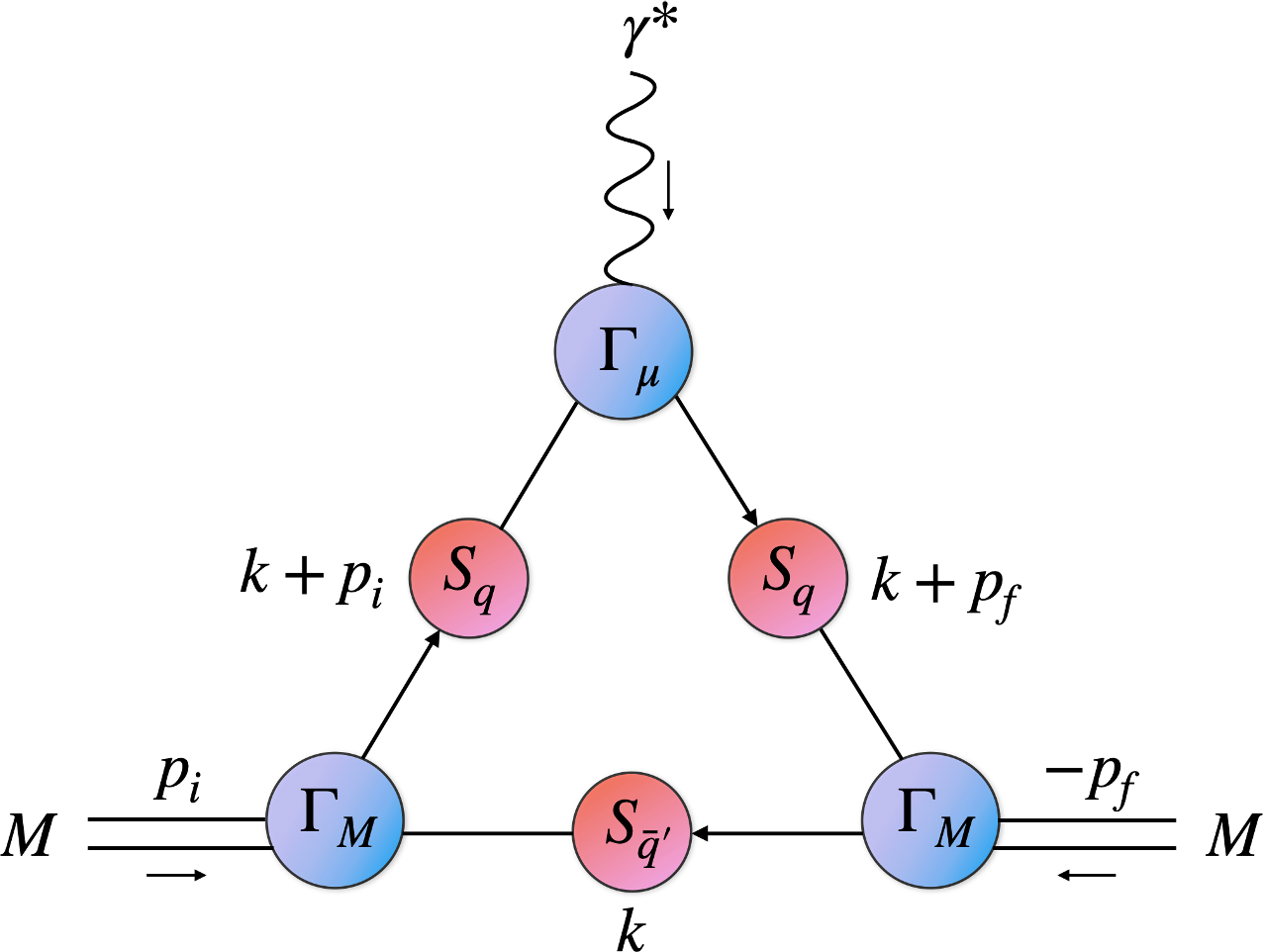}
    \caption{$\M\gamma \M$ vertex, corresponding to the EFF, in the impulse approximation. Here $\Gamma_{\M}$ denotes the $\M$-meson BSA amplitude and $\Gamma_\mu$ the QPV. The edges of the triangle represent the fully dressed quark propagator.}
    \label{fig:TriangleDiagramEFF}
\end{figure}

Considering all the essential ingredients discussed previously, we can now proceed to compute the EFF and TFF in the AM. As we shall see, it turns out to be an analytical task to a large extent. The EFF is written down in the impulse approximation~\cite{Chang:2013nia} which graphically leads to the triangle diagram shown in Fig.~\ref{fig:TriangleDiagramEFF}. Mathematically, this corresponds to the following expression:
\begin{eqnarray}
K_{\mu}F_{\M}^q(Q^2) &=& N_c \, \mbox{tr} \int \frac{d^4k}{(2\pi)^4} \chi_{\mu}^q(k + p_f,k + p_i)   \nonumber \\ 
&\times & \Gamma_{\M}(k_i,p_i) S_{\bar{q}\,'}(k)\Gamma_{\M}(k_f,p_f) \,,
\label{eq:EFFq}
\end{eqnarray}
where $Q$ is the incoming photon momentum and the trace is over spinor indices. The incoming and outgoing meson momenta are denoted by $p_{f,i}=K \pm Q/2$, while the relative momenta of the quark-meson-antiquark vertex are $k_{f,i}=k + p_{f,i}/2$, such that we have $K \cdot Q =0$ and $p_{f(i)}^2=K^2+Q^2/2=-m_M^2$. Note that Eq.~\eqref{eq:EFFq} corresponds to the case when the photon strikes quark $q$. Naturally, we would also have to consider the case when the antiquark is hit. Hence, the total EFF of the meson would be written as:
\begin{eqnarray}
F_{\M}(Q^2) = e_q F_{\M}^{q}(Q^2) + e_{\bar{q}'} F_{\M}^{\bar{q}'}(Q^2) \,,
\label{eq:EFFtot}
\end{eqnarray}
where, $e_{q(\bar{q'})}$ is the quark (antiquark) electric charge, in units of that of the positron\footnote{For neutral mesons composed of same flavored quarks, the total EFF is simply defined as $F_{\M}=F_{\M}^q$.}. 

In order to extract the individual contributions of dressed quarks, $F_{\M}^q$, we contract Eq.~(\ref{eq:EFFq}) with $K^\mu$ and take the corresponding Dirac trace. With the quark propagator, meson BSA and QPV as defined in the previous section, one eventually arrives at:
\begin{eqnarray}
\hspace{-2mm}  F_{\M}^q(Q^2)= \int_{k} \int_{-1}^1 \prod_{i=1}^2 dw_i\rho(w_i)\Lambda_{w_i}^{2\nu_i} \frac{\mathcal{M}_{q,\bar{q}'}(k;K,Q)}{\mathcal{D}_{q,\bar{q}'}^{\nu_{1,2}}(k;K,Q)} \,,
\label{eq:EFF1}
\end{eqnarray}
where the numerator $\mathcal{M}_{q,\bar{q}'}$ has the form,
\begin{eqnarray}
&& \hspace{-10mm} \mathcal{M}_{q,\bar{q}'}(k;K,Q) \equiv \frac{2N_c}{K^2}(4m_M^2 + Q^2)[2k^2 + 2m_q m_{\bar{q}'}]  \nonumber \\
&& \hspace{5mm}  - 2 (K\cdot k) [2k^2 + 4m_q^2m_{\bar{q}'}^2 - 2m_q^2 - 2m_M^2] \,,
\end{eqnarray}
and $\mathcal{D}_{q,\bar{q}'}^{\nu_{1,2}}(k;K,Q) $ contains the product of all denominators related with the quark propagators, BSAs and QPV:
\begin{eqnarray}
&& \hspace{-5mm} \mathcal{D}_{q,\bar{q}'}^{\nu_{1,2}}(k;K,Q) = \Delta(k_{(1+w_1),i}^2, \Lambda_{w_1}^2)^{\nu_1} \Delta(k_{(1-w_2),f}^2 , \Lambda_{w_1}^2)^{\nu_2} \nonumber \\ 
&& \hspace{17mm} \times \; \Delta((k+p_f)^2 , m_q^2) \Delta((k+p_i)^2 , m_q^2) \nonumber \\
&& \hspace{17mm}  \times \; \Delta(k^2 , m_{\bar{q}'}^2) \,,
\end{eqnarray}
with, $k_{(1\pm w),i}=k+(1\pm w)p_i/2$. Additionally, the simplifying notation $\int_{k} \equiv \int \frac{d^4k}{(2\pi)^4}$ has been employed.

To compute the integration on $k$ in Eq.~(\ref{eq:EFF1}), we introduce Feynman parameterization and perform a suitable change of variables. Consequently, we find that the denominator can be simply reduced to $[k^2 + \Omega^2]^{\nu_1+\nu_2+3}$ where $\Omega^2$ has the following form,
\begin{eqnarray}
    \Omega^2 = \frac{1}{4} Q^2 c_0 + \frac{1}{2} (m_q^2-m_{\bar{q}'}^2) c_1 + \frac{1}{4} m_{\M}^2 c_2 + m_{\bar{q}'}^2 \,,
\end{eqnarray}
with $c_{0,1,2}$ given in terms of the Feynman parameters $u_i$ as follows:
\begin{eqnarray}
c_0 &=& [u_1(w_1+1) + 2u_4][u_2(1-w_2) + 2u_3] \,, \nonumber \\
c_1 &=& [u_1(w_1+1) + u_2(w_2+1) + 2u_3 + 2u_4] \,, \nonumber \\
c_2 &=& u_1^2 (w_1+1)^2 + u_2^2 (w_2-1)^2 \nonumber \\
&& + 2u_1(w_1+1) [u_2(1-w_2)+2 u_3 + 2u_4 - 1] \nonumber \\
&& + 2u_2(1-w_2)(2 u_3+2 u_4-1) \nonumber \\
&& + 4[u_3^2 + u_3(2u_4-1) + (u_4-1)u_4] \,.
\end{eqnarray}
Considering that we are computing an elastic scattering, the initial meson is the same as the final meson. Thus, without loss of generality, we choose $\nu=\nu_1=\nu_2$. Hence, the integral in Eq.~(\ref{eq:EFF1}) becomes 
%can be performed using  
%\begin{eqnarray}
%\int \frac{d^4t}{(2\pi)^4} \frac{(t^2)^{\alpha}}{[t^2+a^2]^{\beta}}=\frac{1}{(4\pi)^2} \frac{\Gamma[\beta - \alpha - 2]\Gamma[\alpha +2]}{\Gamma[\beta] \Gamma[2](a^2)^{\beta - \alpha - 2}} \,.
%\label{eq:Gamma´s integral}
%\end{eqnarray}
\begin{eqnarray}
\int_{k} \frac{\mathcal{M}_{q,\bar{q}'}(k,K)}{[t^2+\Omega^2]^{2\nu+3}} &=& \int_{k} \frac{ k^2\mathcal{A} + \mathcal{B} }{[k^2+\Omega^2]^{2\nu+3}} \nonumber \\
&=& \alpha \frac{\mathcal{A}}{[\Omega^2]^{2\nu}} + \beta \frac{\mathcal{B}}{[\Omega^2]^{2\nu +1}} \,, 
\end{eqnarray}
where we have defined:
\begin{eqnarray}
&&  \hspace{-4mm}  \mathcal{A} = 3[u_1(1+w_1) + u_2(1-w_2) + 2u_3 + 2u_4] - 8\,, \nonumber \\
&&  \hspace{-4mm}  \mathcal{B} = \frac{-1}{2} m_{\M}^2 [u_1(1+w_1) + u_2(1-w_2) + 2(u_3+u_4-1)]^2 \nonumber \\ 
&& \hspace{-1mm} \times [u_1(1+w_1) + u_2(1-w_2) + 2(u_3+u_4)] \nonumber \\
&& \hspace{-1mm} + 4 m_q^2m_{\bar{q}'}^2 [u_1(1+w_1) + u_2(1-w_2) + 2(u_3+u_4-1)] \nonumber \\
&& \hspace{-1mm} - \frac{1}{2}Q^2 [u_1(1+w_1)+2u_4][u_2(1-w_2)+2u_3] \nonumber \\
&& \hspace{-1mm} \times [u_1(1+w_1)+u_2(1-w_2)+2(u_3+u_4-2)]    \,,
\end{eqnarray}
with $\alpha$ and $\beta$ being  the following constants
\begin{eqnarray}
    \alpha = \frac{1}{(4\pi)^2} \frac{1}{2\nu(2\nu^2+3\nu+1)} \,,\,\beta = \nu \alpha \,. 
\end{eqnarray}
Thus, the EFF can be expressed as follows
\begin{eqnarray}
F_{\M}^q(Q^2) &=& 2N_c\frac{\Gamma[2\nu+3]}{\Gamma[\nu]^2} \int_{-1}^1 \prod_{i=1}^2 dw_i\rho(w_i)\Lambda_{w_i}^{2\nu} \nonumber \\
&& \hspace{-1.5cm} \times \int_0^{1} du_1 \int_0^{1-u_1} du_2 \cdots \int_0^{1-u_1-u_2-u_3} du_4 \, (u_1 \, u_2)^{\nu-1} \nonumber \\
&& \hspace{-1.5cm} \times \bigg[\frac{\alpha \mathcal{A}}{\Omega^{2(2\nu)}} + \frac{\beta \mathcal{B}}{\Omega^{2(2\nu +1)}} \bigg] \,.
\label{eq:EFF2}
\end{eqnarray}
From the knowledge of the EFF, one can define the flavor contribution to the so called charge radius:
\begin{equation}
    \label{eq:ChargeR}
    (r_{\M}^q)^2 = -\frac{1}{6} \frac{d F_{\M}^q(Q^2)}{d Q^2}\Bigg|_{Q^2 \to 0}\,,
\end{equation}
where the total charge radius is defined in analogy with Eq.~\eqref{eq:EFFtot}. This quantity is important as it is related to the slope of the EFF at zero momentum, and in some cases, it is practically the only reliable information available.

Finally, it is important to recall that the integration on $k$ is performed analytically. The subsequent evaluation of Eq.~\eqref{eq:EFF2} is performed numerically with ease over the entire domain of the Feynman parameters and spectral density $\rho(w_i)$. We now proceed to a similar analysis of the TFFs of pseudoscalar mesons.

%%%%%%%%%%%%%%%%%%%%%%%%%%%%%%%%%%%%%%%%%%%%%%%%%%%%%%%%%%%%%%%%%%%%%%%%%%%%%%%%%%%%%
%%%%%%%%%%%%%%%%%%%%%%%%%%%%%%   Section   %%%%%%%%%%%%%%%%%%%%%%%%%%%%%%%%%%%%%%%%%%
%%%%%%%%%%%%%%%%%%%%%%%%%%%%%%%%%%%%%%%%%%%%%%%%%%%%%%%%%%%%%%%%%%%%%%%%%%%%%%%%%%%%%

\section{Transition form factors}

Complementary to the EFFs under investigation, are the two photon TFFs. Notably, the $\gamma \M\gamma$ interaction vertex is also characterized by a single form factor which is customarily defined as follows:
\begin{eqnarray}
\hspace{-0.26cm} \mathcal{T}_{\mu\nu}(k_1,k_2) &=& \mathcal{T}_{\mu\nu}(k_1,k_2) + \mathcal{T}_{\nu\mu}(k_2,k_1)   \\
 &=& \frac{e^2}{4\pi^2}\epsilon_{\mu \nu \alpha \beta} k_{1\alpha}k_{2\beta}G_{\M}(k_1^2,k_2^2,k_1 \cdot k_2) \,, \nonumber
\end{eqnarray}
where the momentum of the meson is $P=k_1+k_2$, with $k_1$ and $k_2$ the momenta of incoming photons. The case in which one of the photons is on-shell describes the $\gamma\gamma^*\rightarrow \M$ transition, which has been measured experimentally for the $\{\pi,\,\eta,\,\eta',\,\eta_c\}$ systems (see \emph{e.g.} Ref.\,\cite{Danilkin:2019mhd}). Here we focus on the $\{\pi^0,\,\eta_c,\,\eta_b\}$ mesons, employing related experimental and phenomenological information to constrain the parameter space of the AM.
\label{sec:transitionff}
\begin{figure}[!h]
    \centering
    \includegraphics[scale=.31]{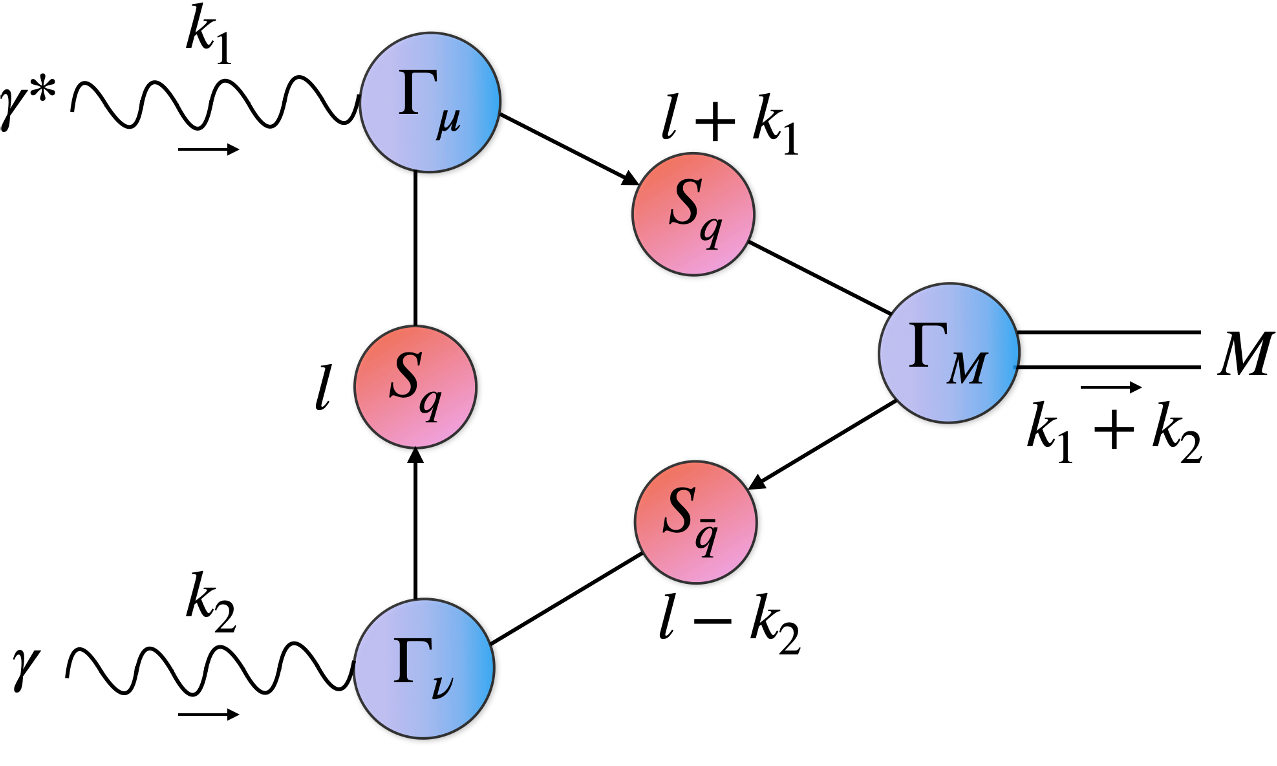}
    \caption{Analogous of Fig.\,\ref{fig:TriangleDiagramEFF} for the TFF.}
    \label{fig:TriangleDiagramTFF}
\end{figure}

In the impulse approximation~\cite{Raya:2015gva}, this form factor is obtained from the triangle diagram depicted in Fig.~\ref{fig:TriangleDiagramTFF}. The standard Feynman rules yield the following mathematical expression:
\begin{eqnarray}
    \mathcal{T}_{\mu\nu}(k_1,k_2)  
    &=& e^2\mathcal{Q}^2_{\M}\mbox{tr} \int_{l} i\chi_{\mu}^q(l,l+k_1)\Gamma_{\M}(l+k_1,l-k_2)  \nonumber \\
    && \hspace{0.3cm} \times S(l-k_2)i \Gamma_{\nu}^q(l-k_2,l) \,,\label{eq:Tmunu}
\end{eqnarray}
where $\mathcal{Q}^2_{\pi,\,\eta_c,\,\eta_b}=\{1/3 ,\,4/9,\,1/9\}$. Moreover, the momenta squared of the virtual and real photons are, respectively, $k_1^2=Q^2$, $k_2^2=0$; thus, the on-shell meson condition imposes $2k_1\cdot k_2=-(m_{\M}^2+Q^2)$.

 Plugging in various quantities from the AM, namely, Eqs.~\eqref{eq:Anzats1}, \eqref{eq:Anzats2} and \eqref{eq:Quark-photon vertex}, into the above expression, the TFF is expressed compactly as follows:
%\begin{eqnarray}
  %G_{\M}(Q^2) &=& \frac{4\pi^2}{e^2}\frac{\epsilon_{\mu \nu \lambda \rho} k_{1\lambda}k_{2\rho}}{\epsilon_{\mu \nu \lambda \rho} k_{1\lambda}k_{2\rho}\epsilon_{\mu \nu \alpha \beta} k_{1\alpha}k_{2\beta}} \nonumber \\ 
 % &\times& \int_{l} \int_{-1}^1 dw \rho(w)\Lambda_{w}^{2\nu}  \mathcal{M}'(l;k_{1,2}) \nonumber \\
%  &\times& \mathcal{D'}(l;k_{1,2}) \;,  
%\end{eqnarray}
\begin{eqnarray}
  G_{\M}(Q^2) &=& \int_{l} \int_{-1}^1 dw \rho(w)\Lambda_{w}^{2\nu}  \frac{\mathcal{M}'(l;k_{1},k_2)}{\mathcal{D'}(l;k_{1},k_2)}\;,  \label{eq:GQfinal}
\end{eqnarray}
where, as in the EFF case, $\mathcal{D'}(l;k_1,k_2)$ denotes the product of denominators from the quark propagator, the BSA and the QPV:
\begin{eqnarray}
    \mathcal{D'}(l;k_1,k_2) &=& \Delta((l-k_2)^2 ,  m_q^2) \Delta((l+k_1)^2 , m_q^2)   \nonumber \\
  &\times& \Delta(l^2 , m_{q}^2)  \Delta(k_{(w-1)}^2, \Lambda_{w}^2)^{\nu} \;.
\end{eqnarray}
Naturally, the scalar function $\mathcal{M}'(l;k_{1},k_2)$ refers to the numerator arising after the Dirac tracing and contraction with a sensible projector operator $\mathcal{P}_{\mu\nu}\sim \epsilon_{\mu \nu \lambda \rho} k_{1\lambda}k_{2\rho}$. Once again, after introducing Feynman parametrization and adequately simplifying the denominator with a suitable change of variables, the TFF can be cast in a form similar to Eq.~\eqref{eq:EFF2}, and the final results can be obtained straightforwardly via numerical integration. 

With semi analytical expressions for the EFF and TFF at hand, we can now proceed to discuss the data-driven method to fix the required model parameters, and analyse the viability of the approach.

%%%%%%%%%%%%%%%%%%%%%%%%%%%%%%%%%%%%%%%%%%%%%%%%%%%%%%%%%%%%%%%%%%%%%%%%%%%%%%%%%%%%%
%%%%%%%%%%%%%%%%%%%%%%%%%%%%%%   Section   %%%%%%%%%%%%%%%%%%%%%%%%%%%%%%%%%%%%%%%%%%
%%%%%%%%%%%%%%%%%%%%%%%%%%%%%%%%%%%%%%%%%%%%%%%%%%%%%%%%%%%%%%%%%%%%%%%%%%%%%%%%%%%%
\section{$\gamma^* \M \to \M$ and $\gamma\gamma^*  \to \M$ Form Factors}
\label{sec:numerics}

In this section we present the collection of results for the EFF ($\gamma^* \M \to \M$) and TFF ($\gamma\gamma^*  \to \M$) of ground-state pseudoscalar mesons $\M= \pi,\,K,\,\eta_c,\,\eta_b$. We start by discussing the data-driven analysis for fixing the 3-4 AM parameters. We then 
compute the predictions of our model
for a wide range of $Q^2$ values. 

 The standard $\chi^2$ statistical test for a set with $N$ number of data points has the expression:
\begin{eqnarray}
    \chi^2=\sum_{i=1}^N \frac{(E_{i}-T_{i})^2}{\delta E_i^2 } \, ,
\end{eqnarray}
where $E_{i}$ represents the $i$-th experimental data to fit, $T_{i}$ is the theoretically predicted $i$-th value for a given point and $\delta E_i$ is the estimated error associated with the $i$-th fitted data point. We implement statistical analysis for the experimental data from the references shown in Table~\ref{tab:fitted_data} to minimize the $\chi^2$.
\begin{table}[h!]
    \centering
    \begin{tabular}{|c|c|c|}
          \hline 
         \, Meson \, & EFF & TFF \\
         \hline \hline 
        $\pi$ & Refs.~\cite{AMENDOLIA1986168,ACKERMANN1978294,JeffersonLabFpi:2000nlc,JeffersonLabFpi-2:2006ysh,JeffersonLab:2008jve} & \, Ref.~\cite{CELLO:1990klc,PhysRevD.57.33,PhysRevD.80.052002,PhysRevD.86.092007}\,  \\
        $K$ & \, Ref. \cite{AMENDOLIA1986435,PhysRevLett.45.232} \, & --- \\
        $\eta_c$ & --- & Ref. \cite{BaBar:2010siw} \\
        %$\eta_b$ & Ref. \cite{} & Ref. \cite{} \\
        \hline
    \end{tabular}
    \caption{Experimental data fitted in our global analysis of EFFs and TFFs of pseudoscalar mesons.}
    \label{tab:fitted_data}
\end{table}
As can be read from the Table\,\ref{tab:fitted_data}, there is more information available for the pion  as compared to other pseudoscalar mesons. For charged kaon, precise data are only available at very low momentum (and, naturally, there is no two-photon to kaon process). There exists data for the $\eta_c$ TFF, which we complement with a theoretical computation of the charge radius,\,\cite{PhysRevC.77.025203}, based upon SDEs. Finally, there is no experimental data concerning the $\eta_b$. So, we rely on the charge radius obtained in Ref.~\cite{Bhagwat2007}, along with the SDE prediction for the TFF from Ref.~\cite{Raya:2016yuj}.

%%%%%%%%%%%%%%%%%%%%%%%%%%%%%%%% Figure %%%%%%%%%%%%%%%%%%%%%%%%%%%%%%%%%%%%%%%%
\begin{figure*}[!t]
    \centering
    \includegraphics[scale=.22]{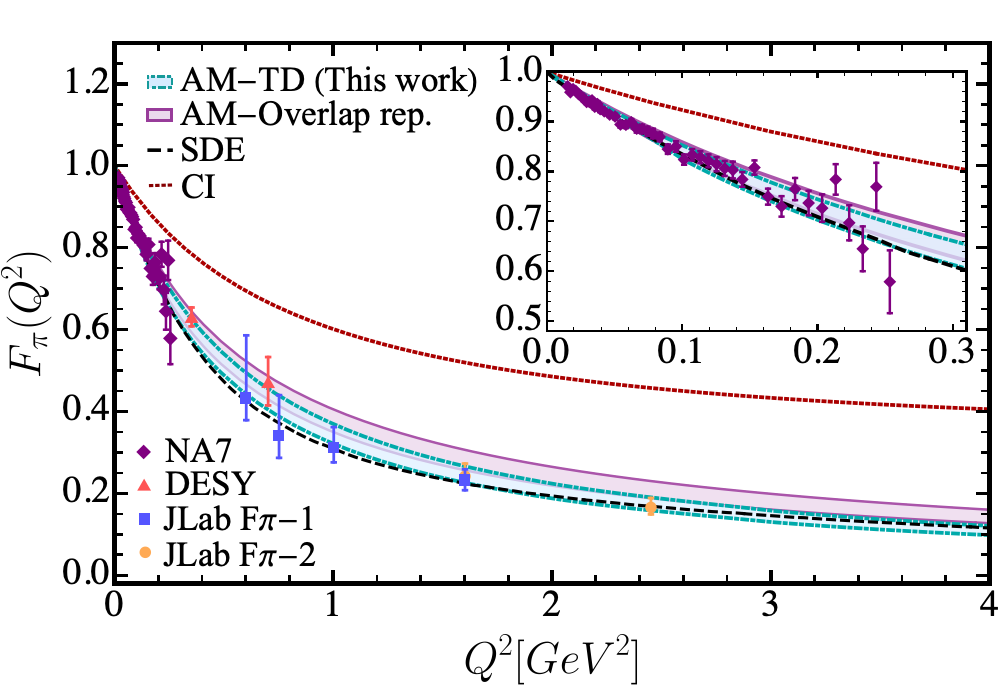}
    \includegraphics[scale=.18]{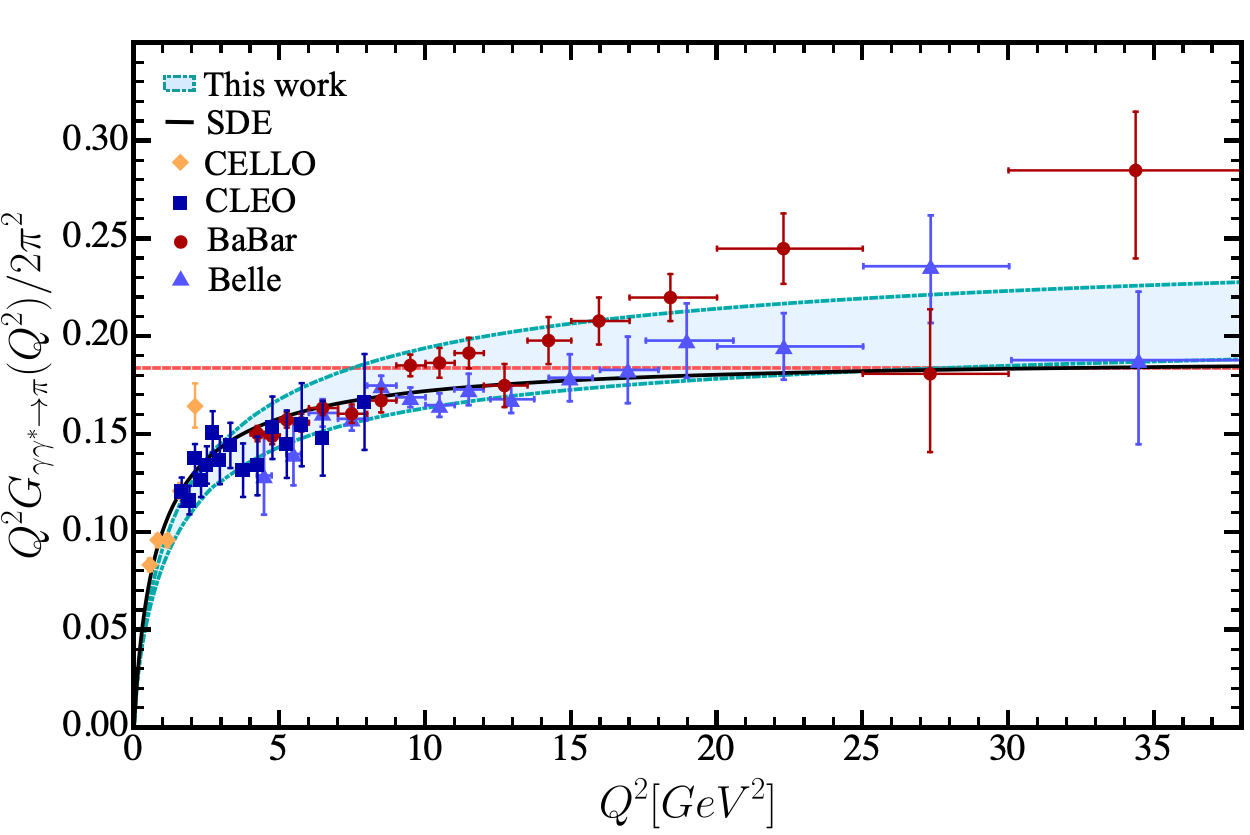}
    \caption{Pion EFF and TFF. Left panel- Pion EFF. Our pion results are represented with the light blue band. The purple band corresponds to the previous AM results from the overlap representation presented in~\cite{Albino:2022gzs}. Dashed black line is the SDE result for the pion~\cite{Chang:2013nia} and the SDE-driven prediction from the CI is represented by the dotted dark red line~\cite{Hernandez-Pinto:2023yin}. Diamonds, rectangles, triangles and circles represent the experimental data from Refs.~\cite{AMENDOLIA1986168,ACKERMANN1978294,JeffersonLabFpi:2000nlc,JeffersonLabFpi-2:2006ysh,JeffersonLab:2008jve}. Right panel- Pion TFF. The blue light band corresponds to our TFFs results of pion. The light red dashed line is the asymptotic limit, $2f_\pi$. The black solid line corresponds to the SDEs results from~\cite{Raya:2015gva}. Experimental results are taken from~\cite{CELLO:1990klc,PhysRevD.57.33,PhysRevD.80.052002,PhysRevD.86.092007}.}
    \label{fig:Pion EFF and TFF}
\end{figure*}

\subsection{$\pi$ meson}
\label{sec:numerics_pion}

We start our analysis with the lightest mesons, namely, pions. 
Besides being  responsible for holding protons and neutrons together in the atomic nucleus, these are the lightest hadrons in nature. Being lightest pseudo-Nambu-Goldstone bosons makes them different from all other hadrons.
Since pions are copiously produced in the high energy collision environments, we take the advantage of the largest amount of data points for both soft and hard processes to constrain the parameters of the AM using available experimental data. We assume isospin symmetry, {\em {i.e.}}, $m_u=m_d$ (implying $m_{\pi^\pm}=m_{\pi^0}$), constraining its value around a typical range of $0.3 \mbox{ GeV} < m_u <0.5 \mbox{ GeV}$. In order to find the best phenomenologically fitted parameters, the pion mass is also allowed to vary in the close vicinity of the value reported in the Particle Data Group (PDG) Ref.~\cite{Workman:2022ynf}. Experimental data from Refs.~\cite{AMENDOLIA1986168,ACKERMANN1978294,JeffersonLabFpi:2000nlc,JeffersonLabFpi-2:2006ysh,JeffersonLab:2008jve,CELLO:1990klc,PhysRevD.57.33,PhysRevD.86.092007,PhysRevD.80.052002} constrain dimensionless parameters $\nu_{\pi}$ and $\alpha_u^{(0)}$ and the dressed quark masses. In Tab.~\ref{tab:pionfit} we present the obtained best parameters to optimize the value of $\chi^2$. 
\begin{table}[h!]
    \centering
    \begin{tabular}{|c|c|c|c|}
         \hline
           $m_u$  & $m_{\pi^\pm}$  & $\nu_{\pi}$ & $\alpha_{u}^{(0)}$  \\
         \hline
           \, 0.3135 \,&\, 0.1395 \, & 0.8428 \,&\, 0.1964 \, \\
         \hline
    \end{tabular}
    \caption{Best fitted values for the pion in the AM. Isospin symmetry is assumed such that $m_u=m_d$. Note that we allow minimal variation in  $m_{\pi^\pm}$  to obtain the best fit. All masses are given in GeV.}
    \label{tab:pionfit}
\end{table}

Results of this exploration are presented in Fig.~\ref{fig:Pion EFF and TFF}. In addition to the agreement with empirical data, the produced curves are highly compatible with sophisticated SDE predictions\,\cite{Chang:2013nia,Eichmann:2019tjk,Raya:2015gva}, and with a previous exploration of the AM employing the overlap representation\,\cite{Albino:2022gzs}. CI model results\,\cite{Hernandez-Pinto:2023yin}, characterized by momentum independent mass functions and BSAs, are also displayed for comparison. From the obtained EFF, one can determine the corresponding charge radius via Eq.~\eqref{eq:ChargeR}, producing:
\begin{align}
    r^{\rm fit}_\pi = 0.67 \mbox{ fm}\,,
\end{align}
which presents a small 1.67 \% deviation from the experimental value $r_\pi^{\rm exp}=0.659(4)\mbox{ fm}$ \cite{Workman:2022ynf}. The error band in Fig.~\ref{fig:Pion EFF and TFF} accounts for a $5\%$ uncertainty of the above value, stemming from a systematic variation of $m_u$.

 In Table~\ref{tab:chi2pion} we report the corresponding $\chi^2$ per set of data points for pions. The global fit has a $\chi^2/d.o.f.\sim 1.93$ confirming quantitatively that the predictions are in agreement with experimental data. Despite the controversy, both BaBar\,\cite{PhysRevD.80.052002} and Belle\,\cite{PhysRevD.86.092007} data sets of the pion TFF were considered in the analysis. The former is mildly disfavored as compared to Belle data; BaBar data set produces $\chi^2/d.o.f. \sim 1.82$, in contrast with the Belle set, from which $\chi^2/d.o.f. \sim 1.45$ is obtained.

 %except for the BaBar data of the TFFs at large $Q^2$. This fact is manifest in Fig. \ref{fig:Pion EFF and TFF}. The error bands reported in Fig. \ref{fig:Pion EFF and TFF} were obtained by allowing a variation of 5\% on the charge radii of the EFF only.

\begin{table}[h!]
    \centering
    \begin{tabular}{|c|c|c|}
    \hline\hline 
          Experiment & \, \# of data points in fit \, & $\chi^2$ \\
         \hline
          NA7\,\cite{AMENDOLIA1986168} & 45 & 48.42 \\
          DESY\,\cite{ACKERMANN1978294} & 2 & 2.50\\
          JLab F$\pi$-1\,\cite{JeffersonLabFpi:2000nlc} & 4 & 1.16\\
          \, JLab F$\pi$-2\,\cite{JeffersonLabFpi-2:2006ysh,JeffersonLab:2008jve}\,  & 4 & 2.56\\
         \hline
          CELLO\,\cite{CELLO:1990klc} & 5 & 83.55\\
          CLEO\,\cite{PhysRevD.57.33} & 15 & 15.44 \\
          BaBar\,\cite{PhysRevD.80.052002} & 17 & 30.95 \\
          Belle\,\cite{PhysRevD.86.092007} & 15 & 21.72 \\
         \hline\hline
          {\bf TOTAL:}& 107 & \,  206.3 \, \\
         \hline\hline
    \end{tabular}
    \caption{Data sets used in our global analysis for pions, the individual $\chi^2$ values, and the total $\chi^2$ of the fit.}
    \label{tab:chi2pion}
\end{table}

It is important to note that the phenomenological agreement is achieved when $\nu_\pi$ tends to unity from below. 
In this context, we recall that QCD prescriptions establish a $1/k^2$ falloff for the dominant BSA of pseudoscalar mesons, which is modified by the corresponding anomalous dimension~\cite{ROBERTS1994477}.  The effects of the latter become apparent in the large momentum regime, but the precise domain in which those manifest strongly depends on the mass sector. Within the AM, such behavior can be emulated by adopting a value of $\nu_{\M}$ close to (but different from) 1. Therefore, this trend is not only acceptable but expected. 
 In contrast with involved numerical evaluations,\emph{e.g.}\,\cite{Raya:2015gva}, a small value for $\alpha^{(0)}_u$ is also favored by data. This means that the correction to the QPV from the transverse terms, given in Eq.~\eqref{eq:alphaq}, although necessary for a proper description of the TFF, is rather mild. Notably, it has been seen that when the quark anomalous magnetic moment is included in a formal manner,~\cite{Dang:2023ysl}, its contribution to the TFF at the $Q^2=0$ limit is commensurate with $\alpha^{(0)}_u$.

\begin{figure}[h!]
    \centering
    \includegraphics[scale=.23]{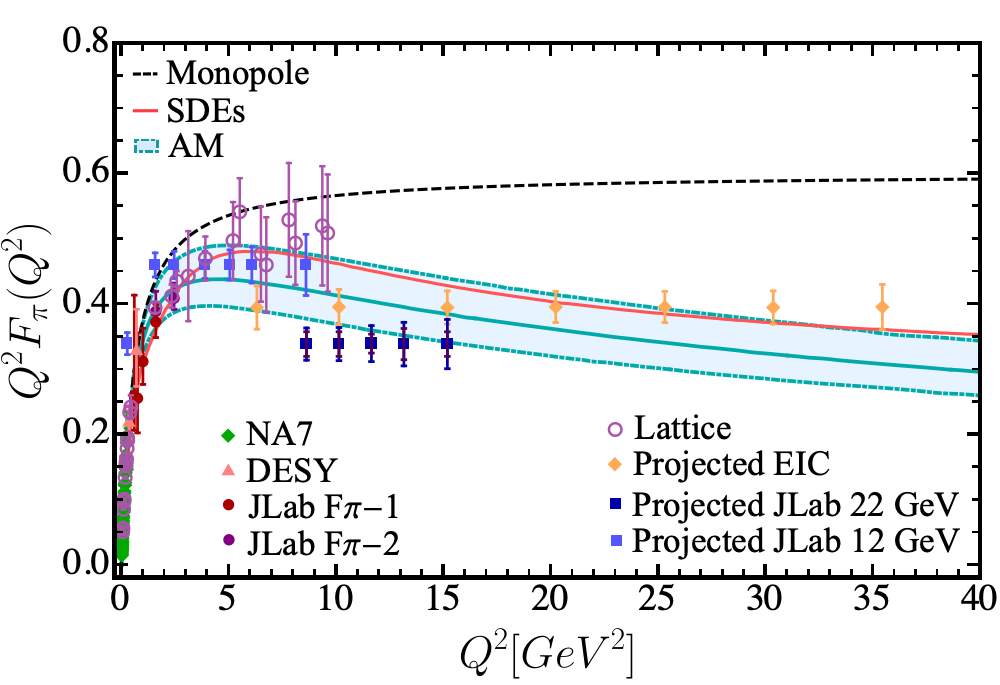}
    \caption{Prediction of the AM for the pion EFF at $Q^2$ up to 40 GeV$^2$. A direct comparison with a monopole fit 
as well as recent lattice results~\cite{Ding:2024lfj} and earlier SDE-based computation from Ref.~\cite{Chang:2013nia} is included.
    Projected EIC and JLab $Q^2$ range~\cite{Arrington:2021biu,Aguilar:2019teb,Accardi:2023chb,Horn:2016rip,Huber:2006} is also depicted. } 
    \label{fig:piEFFEIC}
\end{figure}

We can now analyze the expected theoretical predictions of the AM for EFF of pions for expected center of mass energies of the EIC and JLab. Fig.~\ref{fig:piEFFEIC} displays the EFF in a wide range of $Q^2$ up to $40\text{GeV}^2$. Across the momentum squared domain covered, there is considerable agreement with SDE prediction from Ref.~\cite{Chang:2013nia}. The noticeable contrast of both the calculations with a monopole type fit, supports the fact that scaling violations are already discernible at momentum scales of about $Q^2=4-5\,\text{GeV}^2$,~\cite{Yao:2024drm}. We expect this observation to be confirmed in a convincing manner by the new generation of electron-ion colliders~\cite{Arrington:2021biu,Aguilar:2019teb,Accardi:2023chb,Horn:2016rip,Huber:2006}.

%%%%%%%%%%%%%%%%%%%%%%%%%%%%%%%% Figure %%%%%%%%%%%%%%%%%%%%%%%%%%%%%%%%%%%%%%%%

%%%%%%%%%%%%%%%%%%%%%%%%%%%%%%%% Figure %%%%%%%%%%%%%%%%%%%%%%%%%%%%%%%%%%%%%%%%
\begin{figure*}[t!]
    \centering
    \includegraphics[scale=.22]{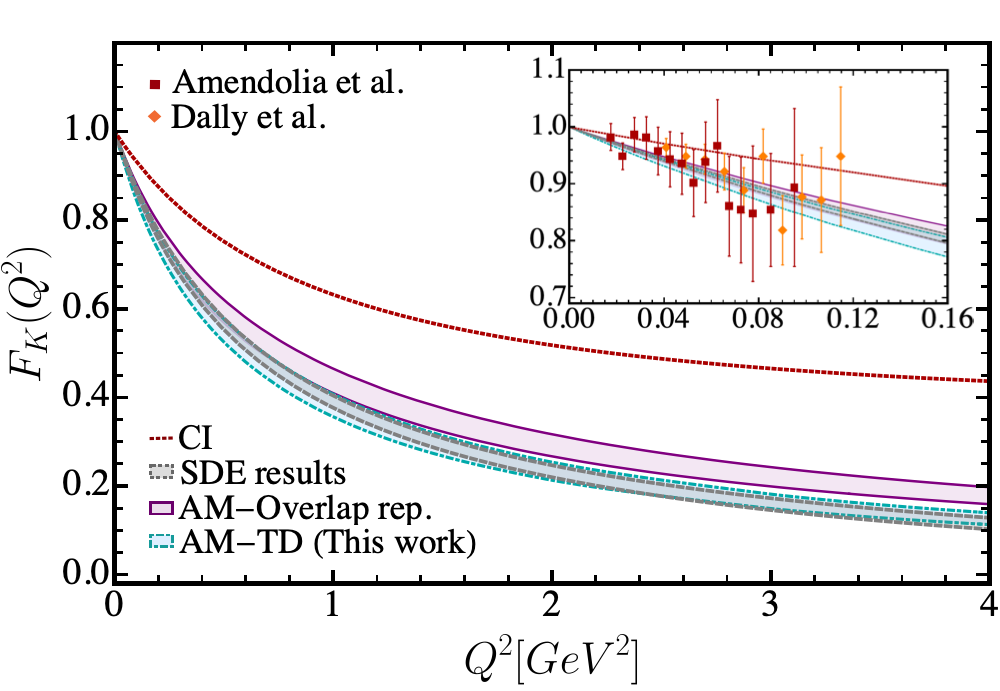}
    \includegraphics[scale=.22]{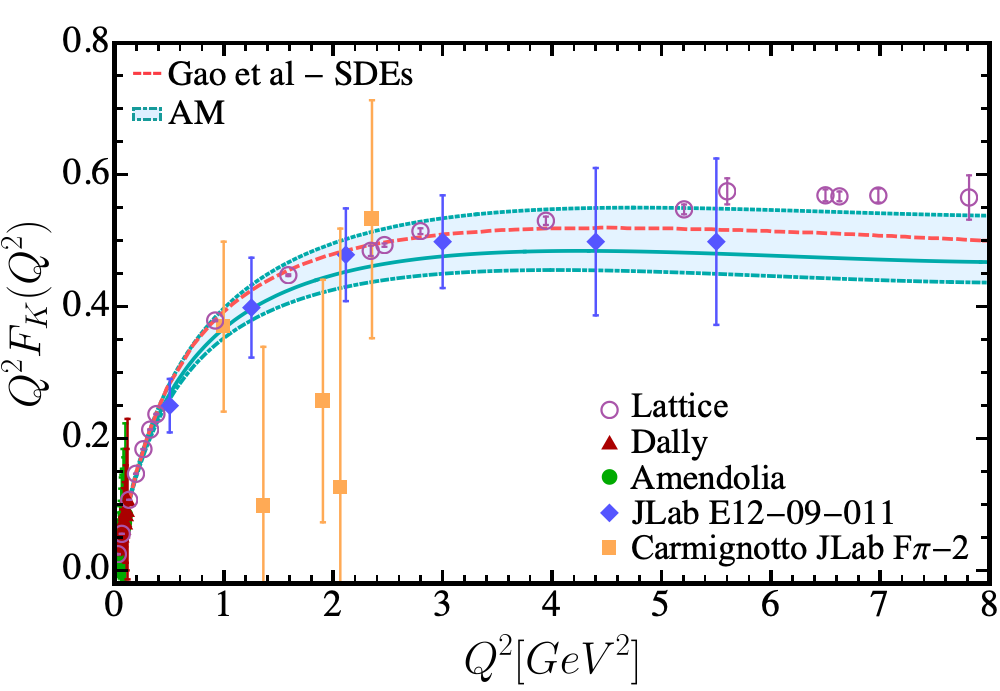}
    \caption{Kaon EFF. Left panel- Kaon $F_K(Q^2)$. The purple band corresponds to the AM previous results from the overlap representation presented in~\cite{Albino:2022gzs}. For comparison, we have included the lower gray band that corresponds to the SDE result for the kaon~\cite{Eichmann:2019bqf,PhysRevD.96.034024} as well as SDE-driven predictions in the CI model (dotted dark red line)~\cite{Hernandez-Pinto:2023yin}. Right panel- $Q^2F_K(Q^2)$ of Kaon. SDE results taken from~\cite{PhysRevD.96.034024} are represented with the dashed light red line. Recent lattice results~\cite{Ding:2024lfj} are also depicted.
In both graphs our results are represented with the light blue band. And, the circles, diamonds, rectangles and triangles represent the experimental data from Refs.~\cite{AMENDOLIA1986435,PhysRevLett.45.232,PhysRevC.97.025204,JLabNew}. }
    \label{fig:Kaon EFFs}
\end{figure*}
%%%%%%%%%%%%%%%%%%%%%%%%%%

\subsection{$K$ meson}
\label{sec:numerics_kaon}

We now turn to the study of kaon pseudoscalar mesons. The understanding of kaons, just like that of pions, is of paramount importance.  Careful scrutiny of the similarities and differences between these two systems is expected to reveal important information regarding the hadronic structure and its connection to the origin of mass in the Standard Model. As explained before, within the present approach, the parameters of the $u/d$-quarks, given in Tab.~\ref{tab:pionfit}, are constrained by the electromagnetic properties of the pion. Similarly, the $s$-quark parameters are fixed by the related empirical data on the kaon. The resulting parameters are shown in Tab.~\ref{tab:kaonfit}.
\begin{table}[h!]
    \centering
    \begin{tabular}{|c|c|c|c|}
         \hline
           $m_s$  & $m_K$ & $\nu_{K}$ & \, $\alpha^{(0)}_{s}$ \, \\
         \hline
           \, 0.5274 \,&\, 0.4936 \, & \, 0.913 \,&\, --  \,\\
         \hline
    \end{tabular}
    \caption{Best fitted values for kaon mesons in the AM. The $u$-quark related values are taken from Tab.\,\ref{tab:pionfit}. Masses are given in GeV. The kaon mass gets fitted to the optimal value given by the PDG~\cite{Workman:2022ynf}.}
    \label{tab:kaonfit}
\end{table}
Large uncertainties on the determination of the experimental data allow us to have $\nu_K = 0.913$ close to (but below) 1. The lack of determination of $\alpha_s^{(0)}$ is due to the fact that its value does not affect the EFF. Therefore, it should be fixed from the two-photon TFF process, which does not exist in this case.

The result for the EFF arising from the fit of the kaon parameters in the AM is displayed in Fig.~\ref{fig:Kaon EFFs}.  The left-hand side of the figure compares our result with that obtained within the SDE approach~\cite{Eichmann:2019bqf}, a previous AM determination via the overlap representation~\cite{Albino:2022gzs}, and the expected harder results of the CI model~\cite{Hernandez-Pinto:2023yin}. The comparison with the data at small $Q^2$, Refs.~\cite{AMENDOLIA1986435,PhysRevLett.45.232}, is also shown.  The EFF has been plotted on a larger span of photon momentum on the right-hand side of Fig.~\ref{fig:Kaon EFFs}. This plot includes comparisons with SDE results~\cite{Gao:2017mmp} and different sets of data~\cite{AMENDOLIA1986435,PhysRevLett.45.232,PhysRevC.97.025204,JLabNew}. Notably, although it was not used in any way to set the parameters, excellent agreement is obtained with the preliminary JLab data~\cite{JLabNew}, which increase the range of $Q^2$ up to 6 GeV$^2$. Just like the case of the pion, the error bands allow for a $5\%$ variation in the kaon charge radius~\footnote{In this case, the variation in $m_u$ produces the band, whereas the ratio $m_s/m_u$ is kept fixed.},
\begin{align}
\label{eq:kaonRad}
    r_K^{\rm fit} = 0.64 \mbox{ fm} \,,
\end{align}
which does not substantially differ from the reported experimental result $r_K\sim 0.56 \mbox{ fm}$~\cite{Workman:2022ynf} and the one obtained in the SDE approach~\cite{Miramontes:2021exi}, $r_K \sim 0.60 \mbox{ fm}$.
In order to be able to quantify the quality of the fit and its predictability, in Tab.~\ref{tab:chi2kaon}, we report the $\chi^2$ corresponding to each set of experimental data on the kaon EFF, producing an overall $\chi^2/d.o.f. \sim $ 0.705 and a rather small $\chi^2/d.o.f. \sim 0.141$ for the preliminary JLab set (not used for the fit).

\begin{table}[h!]
    \centering
    \begin{tabular}{|c|c|c|}
    \hline\hline
          experiment &\, \# data in fit \,  & $\chi^2$ \\
         \hline
          \, Amendolia~\cite{AMENDOLIA1986435} \, & 15 & 5.2822 \\
          Dally~\cite{PhysRevLett.45.232} & 10 &  17.8976 \\
          Carmignotto JLab~\cite{PhysRevC.97.025204} & 5 &  1.3620 \\
          JLab E12-09-011~\cite{JLabNew} & 6 & 0.8444  \\
         \hline\hline
          {\bf TOTAL:}& 36 & 25.3864 \\
         \hline\hline
    \end{tabular}
    \caption{Produced $\chi^2$ values for different available data sets, and its overall value. The first two rows were considered for the fitting procedure of the AM; the next two show the degree of predictability of the model.}
    \label{tab:chi2kaon}
\end{table}

\subsection{$\eta_c$ meson}
\label{sec:numerics_etac}
We now proceed with the analysis of the heavy quarkonia. Note that these systems are characterized by having large dressed quark masses which owe primarily to weak mass generation mechanism of the Higgs field. Including their study is likely to provide productive comparison with their light counterparts, namely pions and kaons. We begin with considering the $\eta_c$ meson. In this case, we have the charm mass $(m_c)$, the $\eta_c$ mass ($m_{\eta_c}$), and the dimensionless parameters $\nu_{\eta_c}$ and $\alpha^{(0)}_c$ to be determined from the phenomenological analysis. 

Due to the fact that the $\eta_c$ is a $q\bar{q}$ system, EFF of $F_{\eta_c}$ is strictly zero (see Eq.~\eqref{eq:EFFtot}) and no experimental input should be expected in this case. However, $F_{\M}^q$ can still be computed theoretically by considering the interaction of the photon only with the quark (or the antiquark). The $\gamma\gamma^*\to \eta_c$ transition, on the other hand, has been measured in BaBar\,\cite{BaBar:2010siw}. We thus employ this phenomenological input for the parameter setting of the AM in this sector. Supplemented by the SDE estimation of the charge radius, $r_{\eta_c}=0.219$\,\cite{Bhagwat2007}, we proceed as before and perform the corresponding $\chi^2$ analysis. The optimal parameters are presented in Tab. \ref{tab:etacfit}. We find similar deviations of $\nu_{\eta_c}$ from unity as in the $\pi-K$ scenario and, in addition, the preferred mass of $\eta_c$ meson deviates from the PDG value~\cite{Workman:2022ynf} by about 4.9\% to obtain best results within our AM. 
\begin{table}[htbp]
    \centering
    \begin{tabular}{|c|c|c|c|}
         \hline
     $m_c$ & $m_{\eta_c}$ & $\nu_{\eta_c}$ & $\alpha^{(0)}_{c}$  \\
         \hline
         \, 1.7364 \,&\, 3.1307 \,&\, 0.8021 \,&\, 0.2669 \,\\
         \hline
    \end{tabular}
    \caption{Best fitted values for $\eta_c$ mesons in the AM. Masses are given in GeV.}
    \label{tab:etacfit}
\end{table}
%%%%%%%%%%%%%%%%%%%%%%%%%%%%%%%% Figure %%%%%%%%%%%%%%%%%%%%%%%%%%%%%%%%%%%%%%%%
\begin{figure*}[t!]
    \centering
    \includegraphics[scale=.22]{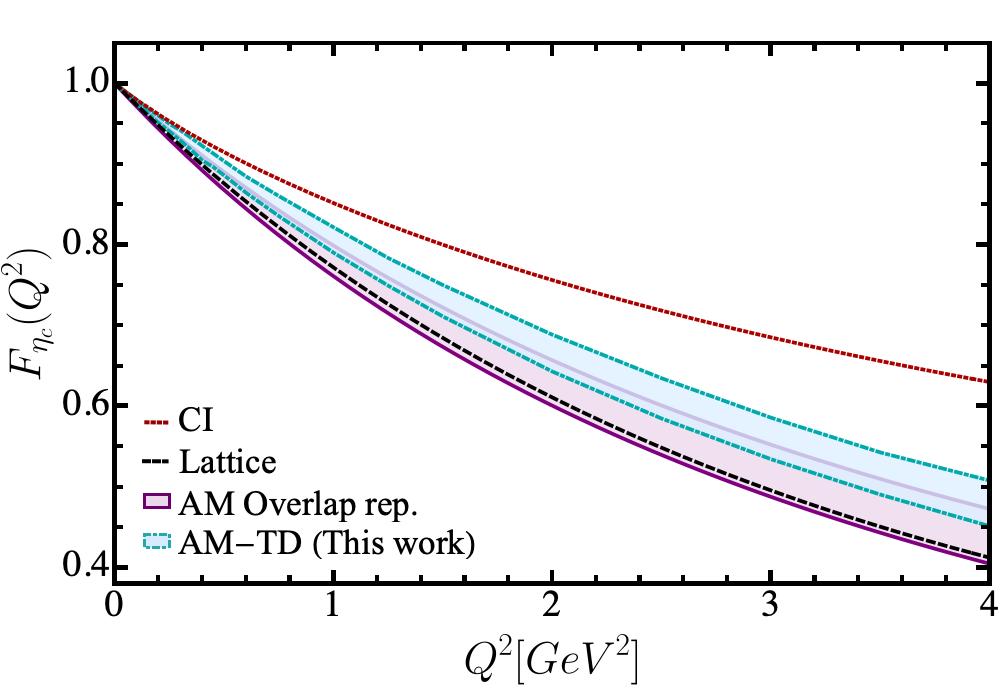}
    \includegraphics[scale=.18]{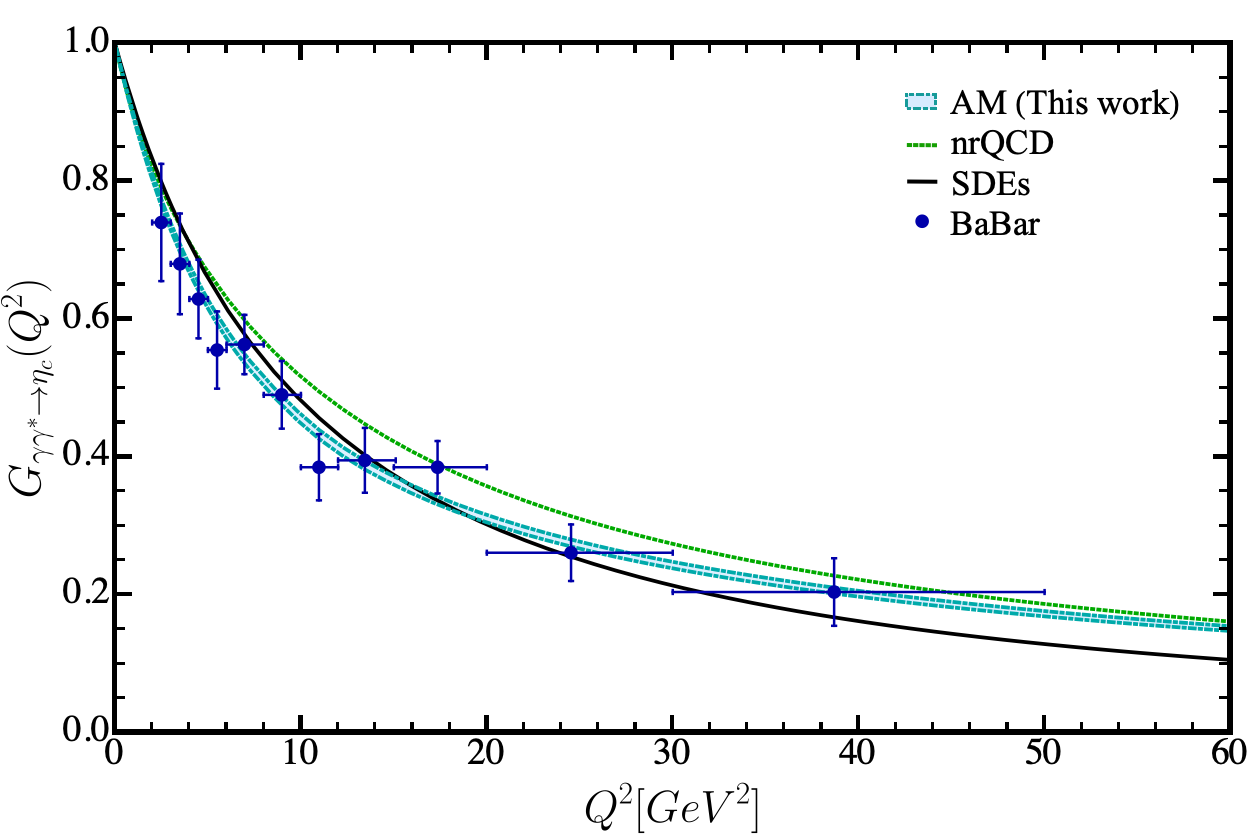}
    \caption{$\eta_c$ EFF and TFF. Left panel- $\eta_c$ EFF. The blue band represents our results with the parameters of Table~\ref{tab:etacfit}. For comparison, we have included the following results: the light purple band that represents the $\eta_c$ previous AM results from the overlap representation discussed in~\cite{Albino:2022gzs}, the dashed black line that corresponds to the lattice QCD results from Refs.~\cite{Dudek:2006ej,Dudek:2007zz}, and the SDE-driven predictions from the the CI model (dotted dark red line)~\cite{Hernandez-Pinto:2023yin}. Right panel- $\eta_c$ TFF. The blue light band corresponds to our TFFs results. The SDEs results~\cite{Raya:2016yuj} are represented by a black solid line, whereas the green dotted curve depicts the NNLO nrQCD predictions~\cite{Wang:2018lry}. The experimental data corresponds to BaBar collaboration from~\cite{BaBar:2010siw}.} 
    \label{fig:Eta_c EFF and TFF}
\end{figure*}
In Tab.~\ref{tab:chi2etac}, we present  the obtained $\chi^2$ of our analysis of $\eta_c$ pseudoscalar mesons. %We point out that lattice phenomenology were sampled over 12 points taken from Ref. \cite{}. These points were taken as pseudo-data for the fit. We assumed errors of 5\% over the central points predicted by lattice. Additionally, 
It is noticeable that in this case $\chi^2/d.o.f\sim 0.36$ which renders the fit almost perfect. With the values tabulated in Tab.~\ref{tab:chi2etac}, we predict a charge radius for $\eta_c$ mesons, $r^{\rm fit}_{\eta_c}$, of
\begin{eqnarray}
    r_{\eta_c}^{\rm fit} = 1.04 \, r_{\eta_c} = 0.228\,\text{fm}  \,.
\end{eqnarray}
This value is also quite compatible with the one computed in lattice QCD, $r_{\eta_c}=0.255(2)~\text{fm}$\,\cite{Dudek:2006ej,Dudek:2007zz}.
%which represents a deviation of XXX \% of the reported value in the PDG \cite{}.

\begin{table}[t!]
    \centering
    \begin{tabular}{|c|c|c|}
    \hline\hline
          \, experiment \, &\, \# data in fit \, & $\chi^2$ \\
         \hline
          $r_{\eta_c}$\cite{Bhagwat2007} & 1 & 0.67 \\
          \hline
          BaBar \cite{BaBar:2010siw} & 11 & 3.97 \\
         \hline\hline
          {\bf TOTAL:}& 12 & \, 4.64 \, \\
         \hline\hline
    \end{tabular}
    \caption{Data sets used in our global analysis for $\eta_c$, the individual $\chi^2$ values, and the total $\chi^2$ of the fit.}
    \label{tab:chi2etac}
\end{table}

In Fig.~\ref{fig:Eta_c EFF and TFF}, we depict the results obtained for the EFF and TFF for the $\eta_c$. On the left hand side, we plot our findings for the $\eta_c$ EFF and compare them with the computed results in lattice QCD~\cite{Dudek:2006ej,Dudek:2007zz}, the CI model\,\cite{Hernandez-Pinto:2023yin}, and the results calculated using the AM within the overlap representation~\cite{Albino:2022gzs}. Just like the pions and the kaons, we again report error bands representing 5\% variation in $r_{\eta_c}$; this same variation is also reported for the TFF depicted on the right-hand side. For the case of the TFF, in addition to the data of the BaBar collaboration,~\cite{BaBar:2010siw},  we also include the SDE prediction~\cite{Raya:2016yuj} and the next-to-next-to leading order (NNLO) non-relativistic QCD (nrQCD) result from~\cite{Wang:2018lry}. As expected, we observe that our results clearly agree with the BaBar measurements and with the theoretical predictions. Once the parameters of the model are carefully fitted, our predictions are readily available for $Q^2$ values as large as experiments can access in the future or lattice QCD as well as other theoretical efforts might be able to produce.

\subsection{$\eta_b$ meson}
\label{sec:numerics_etab}
The heaviest quark found to date in nature which is able to form hadrons is the $b$-quark. It indicates the strongest Yukawa coupling to the Higgs boson, thus providing the 
$b$-quarks with the 
largest mass through explicit chiral symmetry breaking in the Lagrangian. In this section, we study the $b$-quark and the dynamics of the {\em lightest} meson which is composed of it and its antiquark through computing the EFF and the TFF of $\eta_b$. 
%%%%%%%%%%%%%%%%%%%%%%%%%%%%%%%% Figure %%%%%%%%%%%%%%%%%%%%%%%%%%%%%%%%%%%%%%%%
\begin{figure*}[t!]
    \centering
    \includegraphics[scale=.22]{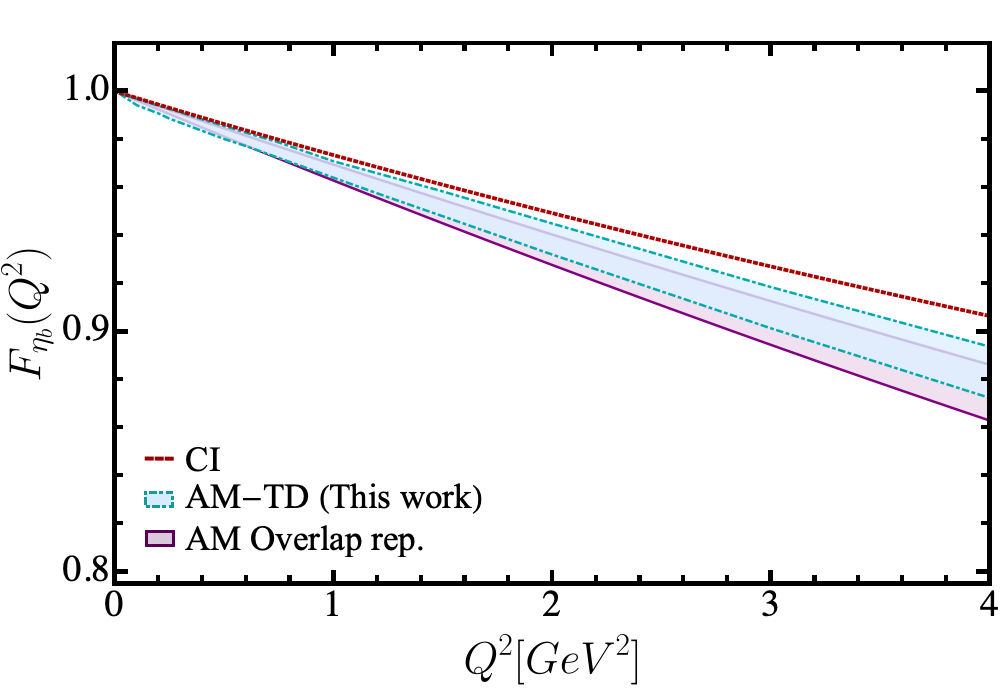}
     \includegraphics[scale=.18]{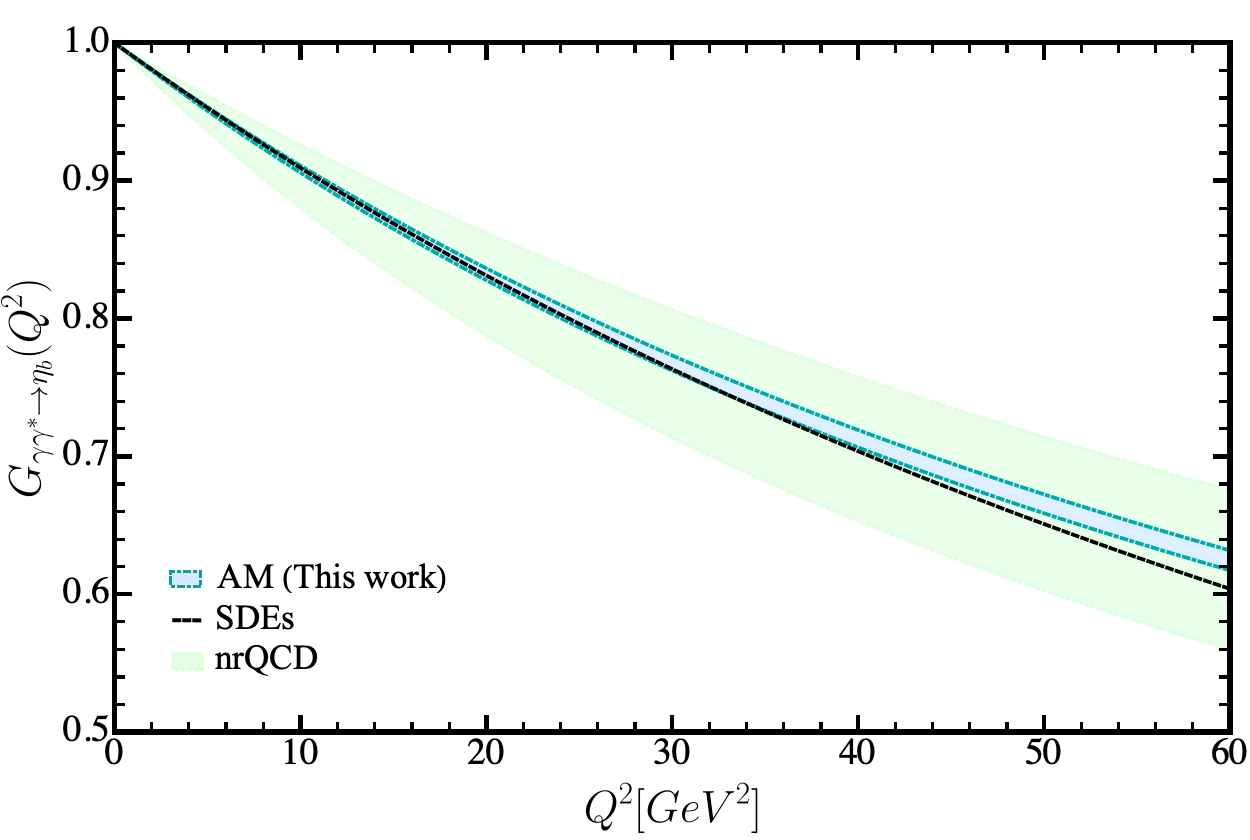}
    \caption{$\eta_b$ EFF and TFF. Left panel- $\eta_b$ EFF. The blue band represents our results with the parameters of Table~\ref{tab:etabfit}. For comparison, we have included the following results: the light purple band that represents the $\eta_b$ previous AM results from the overlap representation~\cite{Albino:2022gzs}; and the SDE-driven predictions proceeding from the CI model (dotted dark red line)~\cite{Hernandez-Pinto:2023yin}. Right panel- $\eta_b$ TFF;    
    the blue light band corresponds to our TFFs results; the black dashed line represents the SDEs results~\cite{Raya:2016yuj}; the green band corresponds to NNLO nrQCD result for $\eta_b$  (the band width expresses the sensitivity to the factorisation scale) from~\cite{Feng:2015uha}.} 
    \label{fig:Eta_b EFF and TFF}
\end{figure*}

\begin{figure*}[t!]
    \centering
    \includegraphics[scale=.22]{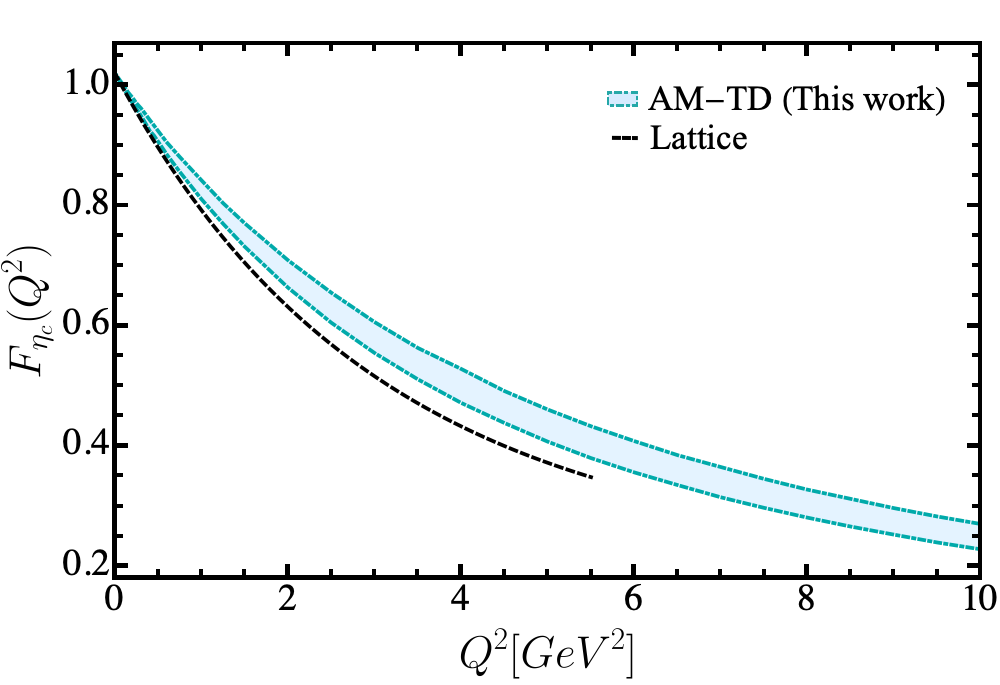}
    \includegraphics[scale=.22]{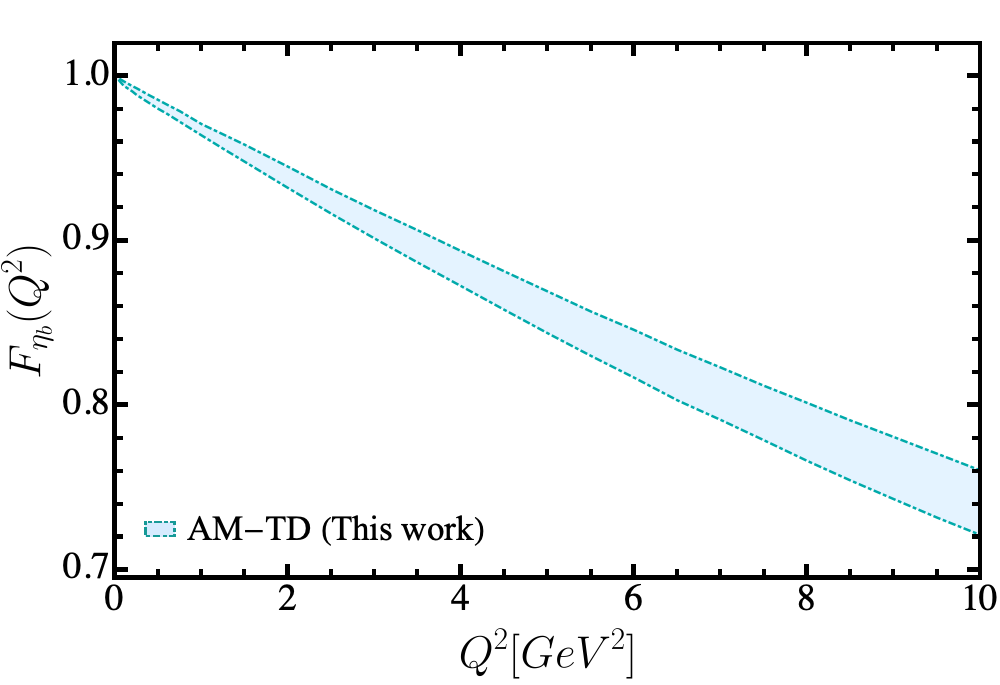}
    \caption{$\eta_c$ and $\eta_b$ EFF at $Q^2$ up to 10 GeV$^2$. Left panel- $\eta_c$ EFF. Right panel- $\eta_b$ EFF. The light blue band corresponds to our EFFs results allowing for a 5\% variation in the charge radius. For comparison, we include the black dashed line on the left plot which corresponds to the $\eta_c$ lattice QCD results from Refs.~\cite{Dudek:2006ej,Dudek:2007zz}. No such computation is available for the $\eta_b$ meson.} 
    \label{fig:Eta_c an Eta_b EFF}
\end{figure*}

Contrary to the previous cases,  there is no analogous experimental information to perform a data-driven global fit for $\eta_b$. Therefore, in order to compute the EFF and the TFF, we proceed with theory-driven phenomenological inputs. We take advantage of the available results for the $\eta_b$ TFF calculated using refined SDE truncations~\cite{Raya:2016yuj}. As all $\gamma^* \gamma \to \{\pi^0,\,\eta,\,\eta',\,\eta_c,\,\eta_b\}$ TFFs have been computed within the same reliable and unified formalism (see Ref.~\cite{Raya:2019dnh}), we obtain agreement with available experiments and QCD constraints. Thereby, the SDE prediction for the $\eta_b$ is expected to be robust. To complement this information,  we also use the $\eta_b$ charge radius $r_{\eta_b}=0.086$ fm, which has been reported in Ref.~\cite{Bhagwat2007}. 

Owing to the lack of data points as such, and instead having a continuous and smooth TFF curve over a large momentum domain (up to $Q^2=60\,\text{GeV}^2$), we construct a random set of points for 12 values of $Q_i^2$, allowing for possible error of $5\%$ with respect to their central values $G_{\M}(Q^2_i)$. Therefore, we can proceed analogously to the other cases. Neither the selection of points nor the size of the error affect the quality of the fit, as long as the selected set covers the entire domain of interest.  The optimal values that minimize the corresponding $\chi^2$ function are collected in Tab.\,\ref{tab:etabfit}. We find a tendency of $\nu_{\eta_b}$ to deviate further from unity, around 27\%, and almost no deviation for $m_{\eta_b}$ as compared to the PDG value, $m_{\eta_b}=9.3987(20)\,\text{GeV}$\,\cite{Workman:2022ynf}. The global analysis in this cases produces a high degree of accuracy, as can be appreciated from the values tabulated in Tab.~\ref{tab:chi2etab}. These numbers produce a pseudo-$\chi^2/d.o.f.\sim 10^{-3}$. 

\begin{table}[h!]
    \centering
    \begin{tabular}{|c|c|c|c|}
         \hline
           $m_b$ & $m_{\eta_b}$ & $\nu_{\eta_b}$ & $\alpha^{(0)}_{b}$  \\
         \hline
           \, 5.3443 \, & \, 9.3886 \, & \, 1.2743 \, & \, 0.1004 \,  \\
         \hline
    \end{tabular}
    \caption{Best fitted values for $\eta_b$ meson in the AM. Masses are in GeV.}
    \label{tab:etabfit}
\end{table}

\begin{table}[h!]
    \centering
    \begin{tabular}{|c|c|c|}
    \hline\hline
          \, experiment \, & \, \# data in fit \, & $\chi^2$ \\
         \hline
          $r_{\eta_b}$\cite{Bhagwat2007} & 1 & 0.006 \\
          \hline
          SDEs \cite{Raya:2016yuj} & 12 & \, 0.004 \, \\
         \hline\hline
          {\bf TOTAL:}& 11 & \, 0.01 \,  \\
         \hline\hline
    \end{tabular}
    \caption{Data sets used in our global analysis for $\eta_b$, the individual $\chi^2$ values, and the total $\chi^2$ of the fit.}
    \label{tab:chi2etab}
\end{table}

Therefore, it is not surprising that there is almost perfect agreement with the model inputs. This level of compatibility is displayed in Fig.~\ref{fig:Eta_b EFF and TFF}, where EFF and TFF of $\eta_b$ are plotted. We reiterate that the reported bands correspond to a variation of $r_{\eta_b}$ of around 5\%. In the case of EFF, we find that
\begin{eqnarray}
    r_{\eta_b}^{\rm fit} = 0.99 \, r_{\eta_b}\,=0.085\,\text{fm}.
\end{eqnarray}
To provide other points of comparison for the resulting EFF, we have included results from the AM in the overlap representation\,\cite{Albino:2022gzs}, as well as those obtained from the CI model\,\cite{Hernandez-Pinto:2023yin}. Regarding the TFF, displayed on the right-hand side of Fig.~\ref{fig:Eta_b EFF and TFF}, SDEs results\,\cite{Raya:2016yuj} and NNLO nrQCD projections\,\cite{Feng:2015uha} are shown. We must emphasize again that our model has the natural advantage that once the small number of model parameters are fixed carefully through the global analysis of experimental data and/or reliable theoretical predictions, our model is designed to provide predictions for arbitrarily large values of the probing photon momentum squared $Q^2$ which is the very aim of every experimental facility and each corresponding theoretical endeavor. 
Fig.~{\ref{fig:Eta_c an Eta_b EFF}} displays our results for the $\eta_c$ and $\eta_b$ EFFs in a larger $Q^2$ domain.

\section{Summary and conclusions}
\label{sec:conclusions}

In this work, we present a detailed calculation of the EFFs and TFFs of the ground-state pseudoscalar mesons $ \pi,\,K,\,\eta_c,$ and $\eta_b$, by considering the interaction vertices $\M\gamma\M$ and $\gamma \M \gamma^*$, respectively, in the impulse approximation. The computation of these form factors is carried out by evaluating the related triangle diagrams which are completely determined by the knowledge of the quark propagator, the meson BSA and the QPV. The results are obtained in an AM developed recently in close connection with QCD.

Within this approach, the structure of the pseudoscalar meson is fully encoded in the BSWF, defined in terms of the quark propagator and the meson BSA. The internal structure of the meson is probed through the electromagnetic QPV. The quark propagator acquires a rather simple form, Eq.~\eqref{eq:Anzats1}, in which the mass function is demoted to play the role a constant dressed quark mass. However, this lack of QCD governed momentum dependence is compensated by providing a suitable {\em Ansatz} for the BSA, expressed as an integral representation, Eq.~\eqref{eq:Anzats2}. Its point-wise behavior is dictated by the so called spectral density. One of the strengths of this approach is that the latter can be extracted systematically and analytically from the knowledge of the meson PDA~\cite{Albino:2022gzs}. Therefore, well-tested realistic understanding of the PDAs is available in the literature which is taken advantage of. This set of distributions and the extraction process is discussed in Appendix~\ref{App}. In addition, through symmetry requirements, a robust but compact expression for the QPV is proposed, Eq.~\eqref{eq:Quark-photon vertex}.

The AM allows different components of the triangle diagrams to be represented in an algebraically convenient way, which leads to writing expressions for the EFF and TFF in a largely analytical manner. In particular, numerical integration over 4-momenta is avoided altogether. Only the integration over compact domains is required, which can be carried out straightforwardly. A comprehensive global analysis is subsequently conducted. Using available experimental data and other phenomenological contributions when needed, we resort to the minimization of the $\chi^2$ function to obtain our best parameters: $m_q$, $m_{q'}$, $m_{\M}$, $\nu_{\M}$ and $\alpha_q^{(0)}$. The final results are presented with added error bands, obtained by allowing a  5\% variation in the charge radii of EFF.

 For all the mesons we obtain a small $\chi^2/d.o.f.$, the largest being 1.93 corresponding to the pion due to the large amount of data available for this meson including the deviation of experimental measurements from the TFFs at large $Q^2$. Besides, in the case of the pion, we observe that our predictions of the EFFs at large energy scales are in agreement with the results of the SDEs and in turn, the AM is capable of mimicking the scaling violations expected for these form factors. This behavior is expected to be corroborated by the projected EIC and JLab experiments. On the other hand, in the case of the kaon, we have that our predictions are again in good agreement with those obtained by the SDEs even at a higher energy scale than those normally reported (unlike the CI model, which produces harder form factors). In turn, they are in almost perfect agreement with the new JLab experimental data which go up to a $Q^2$ of 5.5 GeV$^2$\,\cite{JLabNew}.

Now, regarding heavy quarkonia, the only experimental information available is that related to the $\gamma\gamma^*\to \eta_c$ process, provided by BaBar. Our findings show that such data can be accurately described by the AM. Concerning other theoretical approaches, a measurable difference with the results produced in the CI is found. Notwithstanding, our results are in a perfect agreement with sophisticated SDE treatments and nrQCD. This confluence of results, and the ability to reproduce experimental results, would be a measure of the robustness of the present approach. For the $\eta_b$ meson there is no experimental data available to compare with. However, our theoretical predictions are in complete agreement with those phenomenological predictions obtained within the SDE and nrQCD approaches, as well as with previous determinations of the AM via the overlap representation.

Finally, we conclude that the results generated by the AM compare favorably with all available data, as well as with different theoretical/phenomenological frameworks (including lattice QCD, nrQCD and SDEs). Furthermore, it allows us to extract EFFs and TFFs of pseudoscalar mesons in a large domain of $Q^2$ which will be probed in the upcoming EIC and potential upgrade of the JLab.

%%%%%%%%%%%%%%%%%%%%%%%%%%%%%%%%%%%%%%
%%%%%%%%%%%%%%%%%%%%%%%%%%%%%%%%%%%%%%
%%%%%%    acknowledgements    %%%%%%%%
%%%%%%%%%%%%%%%%%%%%%%%%%%%%%%%%%%%%%%
%%%%%%%%%%%%%%%%%%%%%%%%%%%%%%%%%%%%%%
\begin{acknowledgements}
I.~M.~H-A. thanks {\em Consejo Nacional de Humanidades, Ciencias y Tecnolog\'ias} (CONAHCyT), Mexico, for the scholarship received to carry out her postgraduate studies. The work of R.~J.~H-P is also funded by CONAHCyT through Project No.~320856 ({\em Paradigmas y Controversias de la Ciencia 2022}), {\em Ciencia de Frontera 2021-2042} and {\em Sistema Nacional de Investigadores}. A.~B. acknowledges {\em Coordinaci\'on de la Investigaci\'on Cient\'ifica}, {\em Universidad Michoacana de San Nicol\'as de Hidalgo} grant 4.10. and Fulbright-Garc\'ia Robles scholarship for his stay as a visiting scientist at the Thomas Jefferson National Accelerator Facility, Newport News, Virginia, USA, where part of this work was carried out. A.~B. also thanks Beatriz-Galindo support during his current stay at the University of Huelva, Huelva, Spain. The work of K.~R. is supported by the Spanish MICINN grant PID2022-140440NB-C22, and regional Andalusian project P18-FR-5057. We are grateful to G.~M.~Huber for private communication on the pion and kaon EFF.

\end{acknowledgements}

%%%%%%%%%%%%%%%%%%%%%%%%%%%%%%%%%%%%%%
%%%%%%%%%%%%%%%%%%%%%%%%%%%%%%%%%%%%%%
%%%%%%%%%%%    APPENDIX    %%%%%%%%%%%
%%%%%%%%%%%%%%%%%%%%%%%%%%%%%%%%%%%%%%
%%%%%%%%%%%%%%%%%%%%%%%%%%%%%%%%%%%%%%
\appendix
\section{Spectral density differential equation}
\label{App}
The leading-twist PDA of the pseudoscalar meson, $\phi_{\M}^q(x)$, can be derived through the light-cone projection of the BSWF\,\cite{Chang:2013pq}. Thereby, in the context of the present AM, and models that employ integral representations for the BSWF in general\,\cite{Mezrag:2016hnp,Xu:2018eii,Raya:2021zrz,Albino:2022gzs}, the PDA is expressed in terms of  a meson spectral density $\rho_{\M}(y)$. In our particular case, as detailed in Ref.\,\cite{Albino:2022gzs}, the relationship between $\rho_{\M}(y)$ and $\phi_{\M}^q(x)$ turns out to be well-defined. It is characterized by the following differential equation:
\begin{eqnarray}\label{eq:SpecPDA}
\eta_N \rho_{\M}(w) &=& \lambda_{\nu}^{(2)}(w) \varphi''(w)+\lambda_{\nu}^{(1)}(w) \varphi'(w) \nonumber \\
&+& \lambda_{\nu}^{(0)}(w)\varphi(w)  \,, 
\end{eqnarray}
where $\eta_N$ is the normalization factor for the spectral density, such that $\int_{-1}^1\rho_{\M}(w) dw = 1$. Moreover, we have defined $\varphi(w) \equiv \phi_{\M}^q(\frac{1}{2}(1-w))$. The rest of all the stated quantities are given by the following expressions:
\begin{eqnarray}
\lambda_{\nu}^{(2)}(w) &=& -\frac{1-w^2}{\chi_{+}} \,, \\
\lambda_{\nu}^{(1)}(w) &=& 2 \frac{\nu w}{\chi_{+}} - 2 \frac{\chi_{-}}{\chi_{+}^2} + \frac{\nu \chi_{-}}{\Lambda_{w}^2}\,, \\
\lambda_{\nu}^{(0)}(w) &=& \big\{ 2 \nu \chi_{+}^2 \Lambda_w^2 \left( \chi_{+}^2 - 2 \left( 1 + (1 - \nu) w^2 + \nu \right) \Lambda_w^2 \right)  \nonumber \\
&& \hspace{-1cm} +    4 w (1 - \nu) \left( 2 \Lambda_w^2 - \nu \chi_{+}^2 \right) \Lambda_w^2 \chi_{+} \chi_{-} \nonumber \\
&& \hspace{-1cm} - \left( \nu (1 - \nu) \chi_{+}^4 + 2 \nu \chi_{+}^2 \Lambda_w^2 - 8 \Lambda_w^4 \right) \chi_{-}^2 \big\} / \Theta_w \,, 
\end{eqnarray}
where we identify $\chi_{\pm} \equiv (1-w) M_{\bar{q'}} \pm (1+w) M_{q}$ and $\Theta_w \equiv -4 \left( 1 - w^2 \right) \chi_{+}^3 \Lambda_w^4$. Thus Eq.~\eqref{eq:SpecPDA} entails that the prior knowledge of the PDA determines the spectral density (and vice versa). Now consider the chiral limit ($m_{\M}=0$, $m_{q}=m_{\bar{q}'}$), and $\nu = 1$ (which produces the expected falloff of the BSA in the absence of anomalous dimensions). We then have:
\begin{eqnarray}
\lambda_1^{(2)}=-\frac{(1-y^2)}{2M_q}\;,\;\lambda_1^{(1)}=\lambda_1^{(0)}=0\;,
\end{eqnarray}
which ensures that our AM recovers the known result in~\cite{Chang:2013pq}. In other words, taking 
\begin{eqnarray}
\phi_{asy}(x) = 6x(1-x)  \;,
\end{eqnarray}
we produce 
\begin{eqnarray}
    \rho_{\M}(w)=\rho_{asy}(w):=\frac{3}{4}(1-w^2) \;. 
\end{eqnarray}
For the present study, which focuses on the analysis of $\pi$, $K$, $\eta_c$ and $\eta_b$, the PDAs, we employ the following convenient parameterizations of the PDAs ($\bar{x}=1-x$), reported in Refs.~\cite{Ding:2015rkn}.:
\begin{eqnarray}
\nonumber
\phi_\pi^u(x)&=&20.226\, x\bar{x}\,[1-2.509 \sqrt{x\bar{x}}+2.025x\bar{x}]\;, \vspace{0.2cm} \\ 
\phi_K^u(x)&=&18.04\,x\bar{x}\,[1+5x^{0.032}\bar{x}^{0.024}-5.97x^{0.064}\bar{x}^{0.048}]\;, \nonumber\\
\phi_{\eta_c}^c(x)&=&9.222\, x\bar{x}\,\text{exp}\,[-2.89(1-4x\bar{x})]\;, \nonumber\\
\phi_{\eta_b}^b(x)&=&12.264\,x\bar{x}\,\text{exp}[-6.25(1-4x\bar{x})]\,.\label{eq:PDAsSDE} 
\end{eqnarray}

\vfil\eject

%%%%%%%%%%%%%%%%%%%%%%%%%%%%%%%%%%%%%%%%%%%%%%%%%%%%%%%%%%%%%%%
%%%%%%%%%%%%%%%%%%%%%%%%%%%%%%%%%%%%%%%%%%%%%%%%%%%%%%%%%%%%%%%
%%%%%%%%%%%%%%%%%%%%     BIBLIOGRAPHY      %%%%%%%%%%%%%%%%%%%%
%%%%%%%%%%%%%%%%%%%%%%%%%%%%%%%%%%%%%%%%%%%%%%%%%%%%%%%%%%%%%%%
%%%%%%%%%%%%%%%%%%%%%%%%%%%%%%%%%%%%%%%%%%%%%%%%%%%%%%%%%%%%%%%

%\bibliographystyle{unsrt}
\bibliography{Bibliography}

\end{document}